\documentclass[useAMS, twocolumn, aps, prd, amscd, amsmath, amssymb,verbatim,nofootinbib,usenatbib]{revtex4}
\usepackage{times}
\usepackage{graphicx}
\usepackage{verbatim}
\usepackage{tikz}
\usepackage{color}

\usepackage[T1]{fontenc}
\usepackage{aecompl}
\usepackage{calligra}
\DeclareMathAlphabet{\mathcalligra}{T1}{calligra}{m}{n}
\DeclareFontShape{T1}{calligra}{m}{n}{<->s*[2.2]callig15}{}

\usepackage{hyperref}
\hypersetup{
    pdfnewwindow=true,      % links in new window
    colorlinks=true,       % false: boxed links; true: colored links
    linkcolor=blue,          % color of internal links
    citecolor=green,        % color of links to bibliography
    filecolor=blue,      % color of file links
    urlcolor=blue           % color of external links
}

\def\orb{\text{orb}}

\def\min{\text{min}}
\def\max{\text{max}}

\def\obs{\text{obs}}
\def\em{\text{em}}
\def\GW{\text{GW}}
\def\Msun{{M_{\odot}}}

\def\Mc{\mathcal{M}}

\begin{document}
\title[Repeated Gravitational Lensing of GWs]{Repeated Gravitational Lensing of Gravitational Waves in Hierarchical Black Hole Triples}
\author{Daniel J. D'Orazio}
\email{daniel.dorazio@cfa.harvard.edu}
\affiliation{Department of Astronomy, Harvard University, 60 Garden Street Cambridge, MA 01238, USA}

\author{Abraham Loeb}
\affiliation{Department of Astronomy, Harvard University, 60 Garden Street Cambridge, MA 01238, USA}

\begin{abstract}

We consider binary black holes (BBHs) in a hierarchical triple system where a
more compact, less-massive binary is emitting detectable gravitational waves
(GWs), and the tertiary is a supermassive BH at the center of a nuclear star
cluster. As previous works have shown, the orbital motion of the outer binary
can generate a detectable relativistic Doppler boost of the GWs emitted by the
orbiting inner binary. We show here that for outer-binary orbits with a period
of order one year, there can be a non-negligible probability for repeated
gravitational lensing of the GWs emitted by the inner binary.  Repeating
gravitational lensing events could be detected by the LISA observatory as
periodic GW amplitude spikes before the BBH enters the LIGO band. Such a
detection would confirm the origin of some BBH mergers in nuclear star
clusters. GW lensing also offers new testing grounds for strong gravity.
%
%while non-detection could constrain such formation channels
\end{abstract}

\maketitle

\section{Introduction}

Although gravitational waves (GWs) from merging binary black holes (BBHs) have
been detected \citep{LIGO_O2_COBs:2018}, the astrophysical origin and path
towards merger of these systems remains a mystery.  GWs emitted during
inspiral and merger encode the binary parameters: masses, mass ratios,
eccentricities, and spins. A major goal of the GW-astrophysics community is to
use the measured distributions of these parameters to infer their
astrophysical origins \citep[\textit{e.g.},
Ref.][]{LIGO_O2_BBHproperties:2018}.

For this task, binary parameters are not the only useful information that GWs
can provide. GWs can also carry information on the merger environment. This
can occur either through external actors affecting the GW inspiral, such as
gas dynamics slowing or accelerating the inspiral \citep{YunesKocsisLoeb:2011,
KocsisYunesLoeb:2011, Fedrow+2017, DOrazioLoebSP:2018, Derdzinski+2018}, or
tidal effects in non-BBH inspiral \citep[see Ref.][and references
therin]{LIGOBNSEOS:2018}. The environment of the merger can also affect the
observed waveform without directly affecting the inspiral. Such a situation
arises through frame-dependent effects caused by relative motion of the GW
source or an intervening gravitational potential resulting in relativistic
Doppler boosting and gravitational lensing of the rest-frame GWs, as well as
time-dependent gravitational redshift and the Shapiro-delay
\citep{MeironBenceLoeb:2017}. In this work we consider such 
frame-dependent modulations, focusing on gravitational lensing of GWs.

Previous papers have considered Doppler boost and gravitational lensing of GWs.
The Doppler boost due to peculiar velocities of GW sources has been
investigated for its degeneracy with the source luminosity distance
\citep[\textit{e.g.}, Ref.][]{KocsisStndSiren:2006, Bonvin+2017, ChenLiCao:2019}. Recent
work has also considered the Doppler boost of GWs, detectable in deviations
from the vacuum GW phase, as an indicator of orbital motion around a companion
BH \citep{Inayoshi+Hamian:2017, MeironBenceLoeb:2017,RobsonTriples+2018,
WongBerti+2019} or other intervening mass distributions
\citep{RandallXianyu:2019}.

Lensing of GWs has been considered in the wave and geometrical optics limits
\citep[\textit{e.g.}, Ref.][]{TakaNaka:2003}. The lens itself has been
imagined as an intervening mass at a large distance from the observer and from
the source \citep[\textit{e.g.}, Refs.][]{Ruffa:1999, TakaNaka:2003,
ChristianLoebVitale2018}, but also as a supermassive BH (SMBH) that an
inspiraling BBH orbits in an hierarchical triple \citep{KocsisLens:2013}. The
latter case, most relevant to this study, considers observables in the
magnified GW echoes caused by lensing-induced time delays; these calculations
were carried out for a source that is stationary with respect to the SMBH
lens.

Here we consider for the first time a different lensing regime, the case of a
moving lensed source of GWs. Importantly, we find that such a system can have
orders of magnitude higher probabilities for lensing, exhibit a unique, and
possibly repeating magnification signature in the GW amplitude, and naturally
arise in BBH formation scenarios that involve hierarchical triple systems with
a SMBH tertiary, \textit{e.g.}, Kozai-Lidov oscillation-induced mergers
\citep{AntoniniPerets:2012}, binary-EMRI capture
\citep{ChenHan_bEMRI:2018, ChenLiCao:2019, FernKoby:2018}, and, generally, the
formation of binaries in the disks of Active Galactic Nuclei (AGN)
\citep{Bellovary+2016, StoneAGNI+2017, BartosAGNII+2017, McKernanBBHAGN+2018}.

We estimate that BBH inspirals and mergers that occur while a BBH is on a
year-timescale orbit around a tertiary SMBH could be detected by the Laser
Interferometer Space Antenna \citep[LISA,][]{LISA:2017} as exhibiting strong
(factor of $\sim1.5$) periodic magnification of the GW amplitude. Depending on
astrophysical formation channels, up to of order one percent of all such BBHs
passing through the LISA band could exhibit repeated lensing. Detection
of GW lensing would strongly constrain the distribution of BBHs in galactic
nuclei and offer tests of gravity through GW propagation
\citep{CheslerLoeb:2017}.  Non-detection will help rule out some formation
scenarios.

This work is organized as follows. In \S\ref{S:Lensing Regimes} we present the
probability, timescale, and magnification of GW lensing events in a general
wave-optics treatment, and for a wide range of triple-system parameters. In
\S\ref{S:Lensing Signatures} we illustrate the GW amplitude evolution for two
example lensed systems. In \S\ref{S:Astrophysical Population} we consider the
astrophysical population of such triple systems in order to gauge the
detection probability of a lensed event. In \S\ref{S:Discussion} we discuss
implications for understanding astrophysical formation channels of BBH mergers
as well as studying gravity. In \S\ref{S:Conclusion} we summarize our main
conclusions.

\section{Lensing Regimes}
\label{S:Lensing Regimes}

We begin by specifying the different lensing regimes for hierarchical triples
and the probability for detecting lensing events in these regimes. Throughout
we consider a smaller (in separation and mass) inner BBH, called binary
$i$ with observed total mass $M_i$, orbital frequency $f_i$, and mass ratio
$q_i$. This binary orbits a larger black hole with mass $M_L$ in an
outer binary called binary $o$ with total mass, orbital frequency, and mass
ratio, $M_o$, $f_o$, $q_o$. The chirp mass is given in terms of the binary
total mass as $\Mc = M q^{3/5}/(1+q)^{6/5}$, where $q \leq 1$ is the ratio of
the two binary masses. For a generally eccentric orbit, GWs are emitted by the
inner binary at $n^{\text{th}}$ harmonics of the orbital frequency
$f_{\GW,i}\equiv n f_i$. Throughout we primarily consider binaries on circular orbits where
$n=2$. As the BBH passes behind the tertiary SMBH along the observer's line of
sight, the emitted GWs are lensed. We primarily consider the case where $M_L$
is a larger, SMBH tertiary and the inner binary is one that will
eventually merge in the LIGO band. In all cases we consider only stable
hierarchical triples. 
We gauge stability with the criteria of Ref.
\cite{MardlingAraseth:2001},
\begin{eqnarray} \label{Eq:stab1}
f_i &\geq& q^{1/2}_o Y^{-3/2} f_o \quad \\ \nonumber 
Y &\equiv& 3.3 \left[ \frac{2}{3}\left(\frac{1+q_o}{q_o}\right) \right]^{2/5}  \left(1-0.3\frac{i}{\pi} \right)
\end{eqnarray}
where we conservatively assume that the relative inclination $i$ of inner and outer binary angular momentum is set to $\pi$ radians.

\subsection{Spatial Scales: Wave vs. Geometrical Optics}

Because of the relatively long wavelength of GWs compared to electromagnetic
waves, the wave period can be comparable to the lensing time delay, in which
case we must treat lensing in the wave-optics regime \citep{SEF:Glens:1992}.
The wave- and geometrical optics regimes are delineated by comparing the
time delay and the GW period via the parameter
\citep[\textit{e.g.}, Ref.][]{TakaNaka:2003},
\begin{equation}
\label{Eq:chi}
\chi \equiv  \frac{8 \pi G M_{L} }{c^3} f_{\GW,i},
\end{equation}
where $M_L$ and $f_{\GW,i}$ are the observer-frame (redshifted) quantities.
The inner-binary GW frequency $f_{\GW,i}$ is set by the inner-binary orbital
period and eccentricity. In the limit that $\chi\gg1$, the wave-optics
treatment asymptotes to the geometrical optics case.

While not wholly in the geometrical optics regime, for $\chi\geq1$ lensing
magnification becomes significant. Hence, it is useful to rearrange the above
equation for the frequency above which the transition to geometrical optics
begins and significant magnification is expected,
\begin{equation}
f_{\GW,i} \geq0.08 \text{Hz} \left( \frac{M_L}{10^5 \Msun}\right)^{-1} .
\end{equation}
This means that for GW emitting BBHs orbiting SMBH lenses with masses above
$10^5\Msun$, the transition from wave-optics to geometrical optics begins
before or during the evolution of the BBH through the LISA band ($\sim
10^{-3}$~Hz to $10^{-1}$~Hz for the context of this work).

\subsection{Timescales}

We further delineate three different lensing regimes based upon the motion of
the source with respect to the lens over its observable lifetime. The three
relevant timescales are the orbital period of the outer binary $1/f_o$, the
time in band, $\tau_{\obs}$, and the time for the inner-binary GW source to
cross the Einstein radius of the lens, $\tau_{\text{lens}}$. The resulting
three lensing regimes are
\begin{itemize}
\item The Repeating-Lens Regime: $1/f_o \lesssim \tau_{\obs}$
\item The Slowly-Moving Lens Regime: $1/f_o \geq \tau_{\obs} \geq \tau_{\text{lens}}$
\item The Stationary-Lens Regime: $1/f_o \geq \tau_{\obs} \leq \tau_{\text{lens}}$.
\end{itemize}
The time in band $\tau_{\obs}$ is set by the time that the inner binary emits
GWs above a set signal-to-noise ratio (SNR) in all GW-bands through which the
binary passes over its lifetime, and $\tau_{\text{lens}} \equiv
r^{\max}_E/v_{\orb, o}$, where $v_{\orb, o}$ is the orbital velocity of the
inner binary around the tertiary SMBH, and $r^{\max}_E = \sqrt{4 G M_L/c^2 a_o
\sin{I_o}}$ is the Einstein radius when the inner binary (source) is directly
behind the tertiary (lens), for outer binary inclination $I_o$ \citep[see
\textit{e.g.}, Ref.][]{DoDi:2018}.

In the repeating-lens regime we require, as a hard limit on detection of such
a flare, that the timescale constraint $\tau_{\text{lens}} \gg f^{-1}_{\GW,i}$ be
satisfied. We find this only to affect very short outer-orbital periods for
small tertiary masses, which we have already shown exhibit low magnification
and, as we show in the subsequent sections are low probability events due to rapid
decay of the inner binary into the tertiary SMBH.

\subsection{Probabilities}

The probability for the tertiary SMBH to lens GWs emitted by the inner binary
can be computed separately in each of the above-mentioned regimes.
Conservatively, we focus on the geometrical optics limit. In the wave-optics
limits, these probabilities are higher \citep{Ruffa:1999}, but as we discuss
below, the magnification can be lower in that case, and not as observationally
interesting. Hence, we compute the probability of a significant lensing event
in the geometrical optics limit as in Refs. \cite{DoDi:2018, DoDi:2019}. To do
so we define a significant lensing event as one in which the projected
distance between the GW source and the SMBH lens falls within one Einstein
radius (see $r^{\max}_E$ above). The lensing probability is then computed from
the geometrical cross section of the Einstein radius compared to the total
area available to the inner binary along its orbit around the tertiary, $M_L$:
\begin{enumerate}
\item In the repeating-lens regime we observe the system for longer than an orbit of the outer binary and so treat the outer orbit as a ring that can be randomly oriented with respect to the line of sight. Then the probability that this ring falls within one Einstein radius of the lens is the lensing probability in the repeating-lens regime,
	\begin{eqnarray}
	\mathcal{P}_{\text{rl}} &\approx& \frac{2}{\pi} \sin^{-1}\left( \frac{r^{\max}_E}{a} \right) \\ \nonumber 
	&=& \frac{2}{\pi} \sin^{-1}\left[ \sqrt{\frac{4 G M_L}{c^2}}  \frac{ \left(2 \pi f_o \right)^{1/3}}{\left[G M_{o}\right]^{1/6}} \right] .
	\end{eqnarray}
	This is the probability that the source, observed for one orbit, will be
	inclined within one Einstein radius of the lens. This condition sets the
	required outer-binary inclination given the outer-binary separation and
	the lens mass \citep[see Eq. (6) of Ref.][]{DoDi:2018}. The duration of strongly
	lensed emission in this regime is given approximately by
	$0.5\mathcal{P}_{rl} P_o$.
\item  
	In the slowly-moving regime, the outer binary does not execute a full orbit while GWs from the inner binary are detectable, and the above probability is simply decreased by the fraction of an outer orbit that is observed.
	\begin{equation}
	\mathcal{P}_{\text{sm}} \approx \min\left\{\frac{\tau_{\obs} f_o}{2}, 1\right\} \mathcal{P}_{rl} .
	\end{equation}
\item In the stationary-lens regime the probability decreases to the ratio of the area enclosed by the Einstein radius to the sphere of possible positions of the inner binary on its orbit at an arbitrary inclination around the lens,
	\begin{eqnarray}
	\mathcal{P}_{\text{sl}} &\approx& \frac{GM_L}{a_o c^2} = \dfrac{ \dfrac{G M_L}{c^3} 2\pi f_o}{\left(\dfrac{GM_{o}}{c^3}2\pi f_o\right)^{1/3}} \nonumber \\
	&\approx& \left(\frac{GM_{o}}{c^3}2\pi f_o\right)^{2/3}.
	\end{eqnarray}
\end{enumerate}
We label the general probability of a lensing event, encompassing all of the
above lensing regimes, $\mathcal{P}$.

%%%%%%%%%%%%%%%%%%%%%%%%%%%%%%%%%%%%%%%%%%%%%%%%
%%% FIGURE %%%
%%%%%%%%%%%%%%%%%%%%%%%%%%%%%%%%%%%%%%%%%%%%%%%% 

\begin{figure}
\begin{center}%$
\includegraphics[scale=0.27]{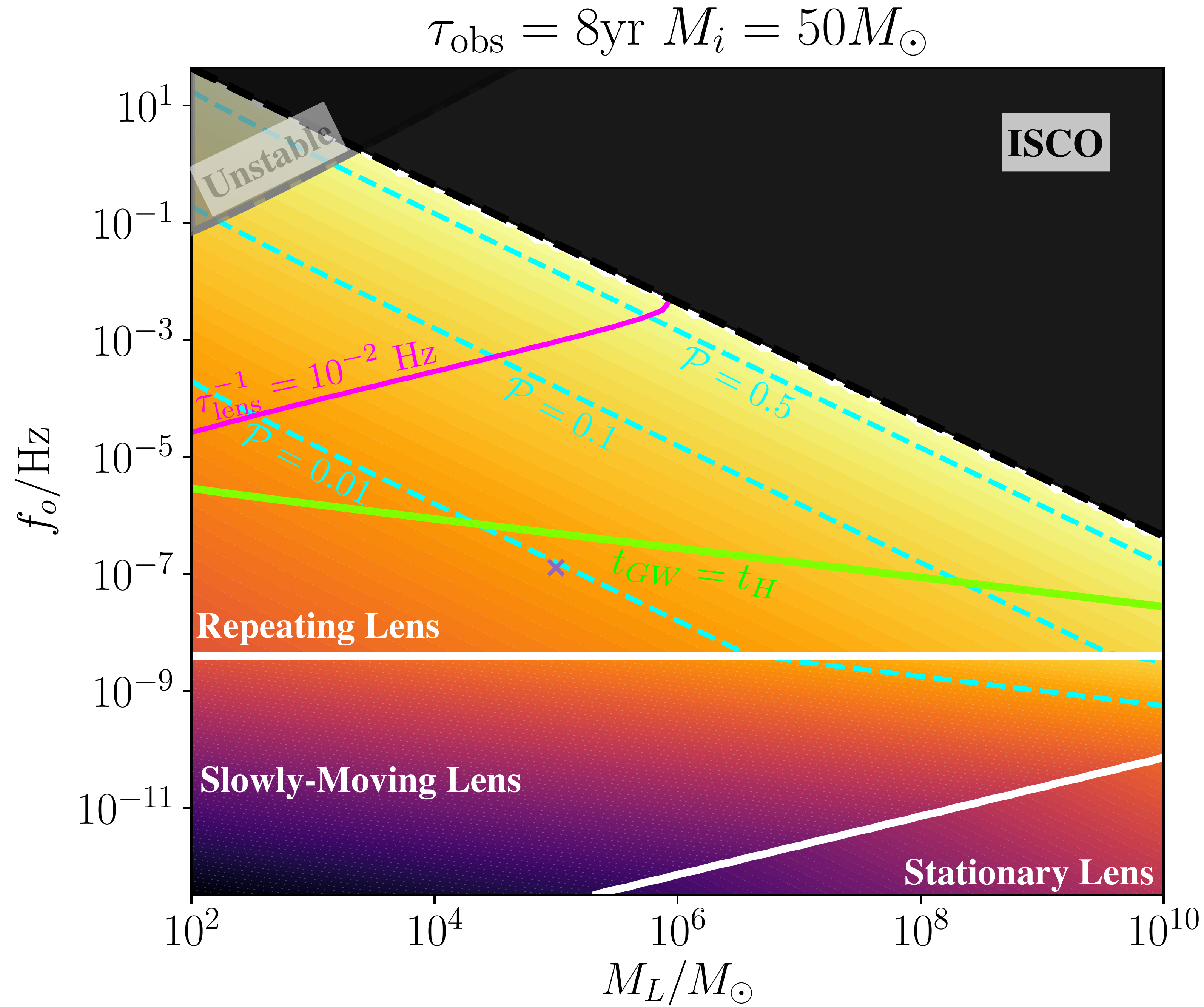} %&
\end{center}
\vspace{-10pt}
\caption{
Contours of the lensing probability $\mathcal{P}$ delineated into the three
lensing regimes discussed in the text (marked by the solid white lines). The
vertical axis denotes the orbital frequency of the outer binary  and the
horizontal axis denotes the mass of the tertiary BH that the inspiraling BBH
orbits. The cyan dashed lines pick out specific lensing-probability contours.
The black region denotes unphysical outer orbits set by the innermost stable
circular orbit (ISCO) and the gray region denotes unstable hierarchical
triples (assuming the inner binary is separated such that it will merge in a Hubble time). Above the bright green line, the outer binary inspirals in less than
a Hubble time, and hence, as discussed in  \S\ref{S:Astrophysical Population},
such systems may be less common. Triple systems falling above the magenta line
will have lensing durations that are shorter than than 100 seconds, and are,
hence, not detectable at the ($\sim 10^{-2}$~Hz) GW frequency band of
interest.
}
\label{Fig:tscales}
\end{figure}

Figure \ref{Fig:tscales} shows contours of these lensing probabilities for a
range of relevant outer binary frequencies (vertical axis) and lens masses
(horizontal axis). Here purple contours denote low probabilities while yellow contours
denote high probabilities, with the labeled dashed-cyan lines delineating
$1\%$, $10\%$, and $50\%$ probabilities. These probabilities are computed
assuming that the emitting BBH falls within one Einstein radius of the lens in
projection, for randomly oriented outer orbits. The black and gray regions
shade forbidden orbits due to the innermost stable circular orbit (ISCO) of
the outer binary, and triple stability (Eq. \ref{Eq:stab1}), respectively.

The white labeled lines are drawn to delineate the three different timescale
regimes discussed above. It is apparent that the highest probabilities for
lensing occur for short outer orbits, in the repeating-lens regime, while
chance lensing of BBHs on longer period orbits are rare. As we will discuss in
the next section, however, shorter outer orbits may be depleted by rapid
orbital decay into the central SMBH. In green we have drawn a line where the
GW-induced inspiral time of the BBH into the SMBH is equal to the Hubble time,
for a circular outer orbit. Systems falling above this green line are less
likely to last long enough to fulfill their full lensing probability. The
magenta line denotes systems above which the lensing timescale is shorter than
the period of a GW at $10^{-2}$~Hz. The majority of triple
parameter space allows lensing flares that are long enough in duration for
detection in the LISA band.

\vspace{10pt}

To draw Figure \ref{Fig:tscales}, we have chosen an inner-binary with equal
mass components, a total mass of $50\Msun$, and a separation such that it will
merge in a Hubble time. We choose an observation time of $8$~yrs
corresponding to a maximum observation time in the LISA band
\citep[\textit{e.g.}, Ref.][]{SesanaMultibandGW:2016}. For longer (shorter)
observation times, the white lines delineating the different lensing regimes
shift downwards (upwards). Hence, this plot is qualitatively similar for year
timescale observations that would apply to BBH inspirals that leave the LISA
band during the LISA mission, or for those observed in a 4-year mission. For
the heavy stellar-mass BBHs that can be observed for year timescales with
LISA, the most promising lensing systems are those with a $\sim0.1-8$~year
outer-orbital period and a central SMBH with mass in the range of
$10^5\Msun-10^{10}\Msun$. As discussed above, the lensing magnification will
be low in the LISA band for SMBH lenses below $10^5 \Msun$.

\subsection{Magnification}

In the case of a point mass lens, the GW-amplitude magnification factor due to lensing can be written in terms of the parameter $\chi$ \citep{TakaNaka:2003}, defined in Eq. (\ref{Eq:chi}),
\begin{widetext}
\begin{eqnarray}
\label{Eq:Mag1}
F(\chi) &=& \exp\left\{ \frac{\pi^2 \chi}{2} + \sqrt{-1} \pi \chi \left[ \log\left(\pi \chi\right) - 2 \phi_m(u) \right] \right\} \Gamma\left(1 - \sqrt{-1}\pi \chi \right) \ {}_1F_1\left(\sqrt{-1} \pi \chi,1; \sqrt{-1} \pi \chi u^2 \right) \\ \nonumber
\phi_m(u) &\equiv& \left(x_m - u \right)^2/2 - \log x_m, \qquad x_m = \frac{u+\sqrt{u^2+4}}{2}
\end{eqnarray}
\end{widetext}
where $u$ is the separation between source and lens in units of the Einstein
radius,  $r_E = \sqrt{4 G M_L/c^2 a_o \sin{I_o} \sin\left( 2 \pi/P_o t \right) }$, assuming a circular orbit here and for outer
binary inclination $I_o$ \citep[see \textit{e.g.}, Ref.][]{DoDi:2018}, and
${}_1F_1$ is the confluent hypergeometric function of the first kind.

For $\chi\ll 1$ the GWs are diffracted by the BH, resulting in very little
magnification. As $\chi$ approaches unity, however, GWs traversing the lens
BH, being coherent, interfere resulting in a time-oscillatory magnification
with amplitude approaching that of the point mass geometrical optics limit
($|\mu_+| + |\mu_-|$) and oscillation frequency given by the lens time delay
$\tau_d$ and the GW frequency,
\begin{equation}
|F|^2 \rightarrow |\mu_+| + |\mu_-| +  2|\mu_+ \mu_-|^{1/2} \sin{\left( 2 \pi f_{\GW,i} \tau_d \right)} ,
\label{Eq:Mag2}
\end{equation}
where $\mu_{\pm} = 1/2 \pm (u^2+2)/(2u\sqrt{u^2+4})$. The time delay between the two lensing `images' is given by
\begin{equation}
\tau_d = 4 M_L \left[ \frac{u}{2}\sqrt{u^2 +4} + \ln{ \left(\frac{\sqrt{u^2+4} +u}{\sqrt{u^2+4} -u} \right)} \right] ,
\label{Eq:Magtd}
\end{equation}
where again, $M_L$ is the redshifted quantity. In the next section we
explore magnification signatures in the GW amplitude using the general form of the magnification factor, Eq. (\ref{Eq:Mag1}).

\section{Lensing Signatures}
\label{S:Lensing Signatures}

%%%%%%%%%%%%%%%%%%%%%%%%%%%%%%%%%%%%%%%%%%%%%%%%
%%% FIGURE %%%
%%%%%%%%%%%%%%%%%%%%%%%%%%%%%%%%%%%%%%%%%%%%%%%%    
\begin{figure*}
\begin{center}$
\begin{array}{cccc}
\includegraphics[scale=0.55]{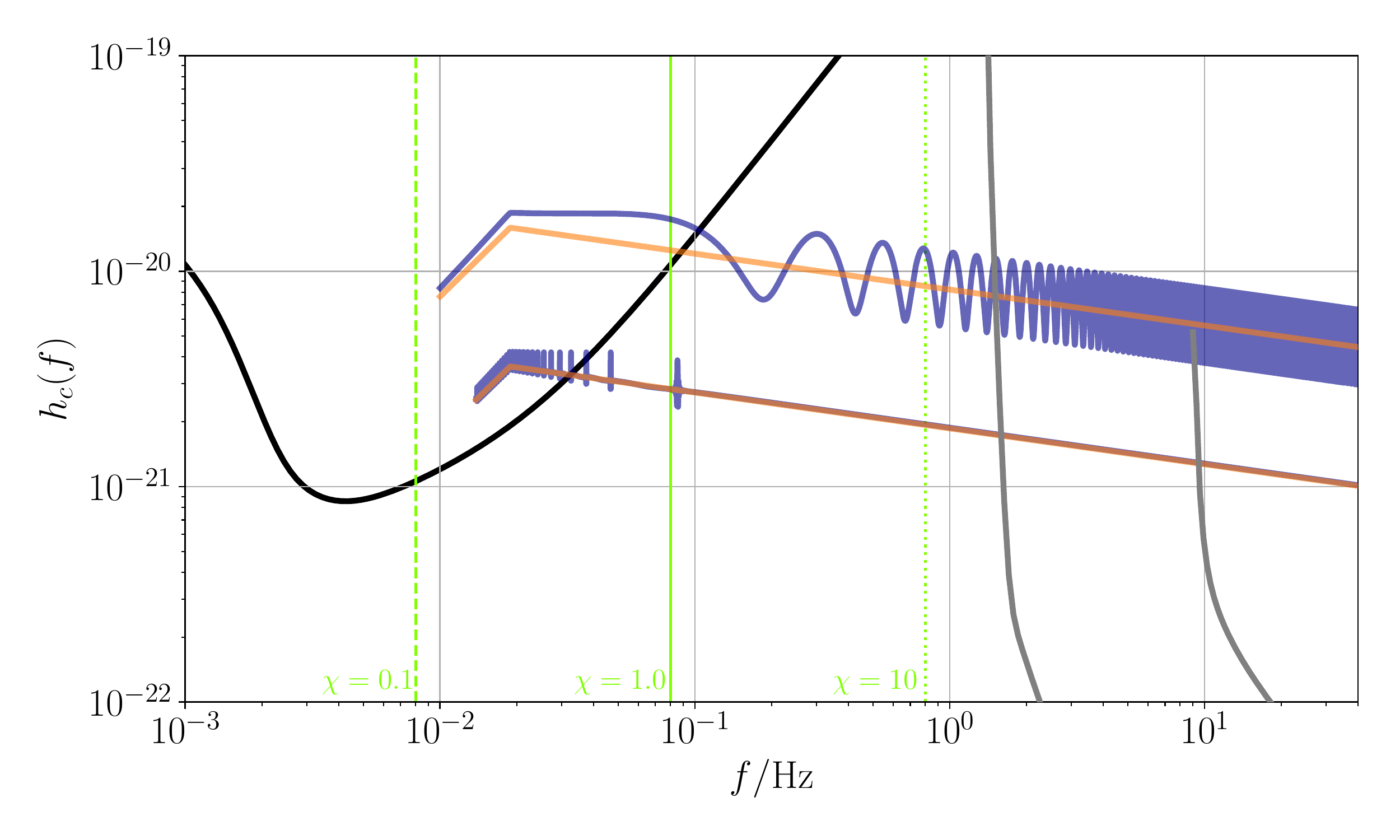}
\end{array}$
\end{center}
\vspace{-20pt}
\caption{
Example lensing strain tracks for a $50 \Msun$ BBH with equal mass components orbiting a $10^5 \Msun$ SMBH (lens). The orange lines represent the unlensed characteristic strain amplitude of the BBH inspiral and purple curves represent the lensed characteristic strain amplitudes. The upper purple track, which oscillates in amplitude as it leaves the LISA band, is drawn for the (much less probable) stationary-lens case, where the merging BBH is on a wide orbit around the central SMBH and has a projected separation from the SMBH lens of one Einstein radius ($u=1$). The lower amplitude track is for a BBH on a $0.25$~year orbit that is lensed 32 times over the 8 years until merger considered here. The closest approach of the inspiraling BBH source and the SMBH lens is one Einstein radius ($u=1$), at the peak of the lensing flares. Green vertical lines denote where the geometric factor $\chi$ is equal to 0.1, 1, and 10, with $\chi\leq 1$ denoting the wave-optics lensing regime. The two inspirals are placed at redshifts $z=0.05$ and $z=0.2$ for ease of visualization. The black line is the LISA sensitivity curve from \cite{RobsonLISASens:2019}, while the gray curves represent the sensitivities of LIGO O2 and a planned third generation detector \cite[the Einstein Telescope][]{ET}. The repeating-lens case is explored further in Figure \ref{Fig:zoom}.
}
\label{Fig:hcflens}
\end{figure*}

%%%%%%%%%%%%%%%%%%%%%%%%%%%%%%%%%%%%%%%%%%%%%%%%
%%% FIGURE %%%
%%%%%%%%%%%%%%%%%%%%%%%%%%%%%%%%%%%%%%%%%%%%%%%%    
\begin{figure*}
\begin{center}$
\begin{array}{cc}
\includegraphics[scale=0.35]{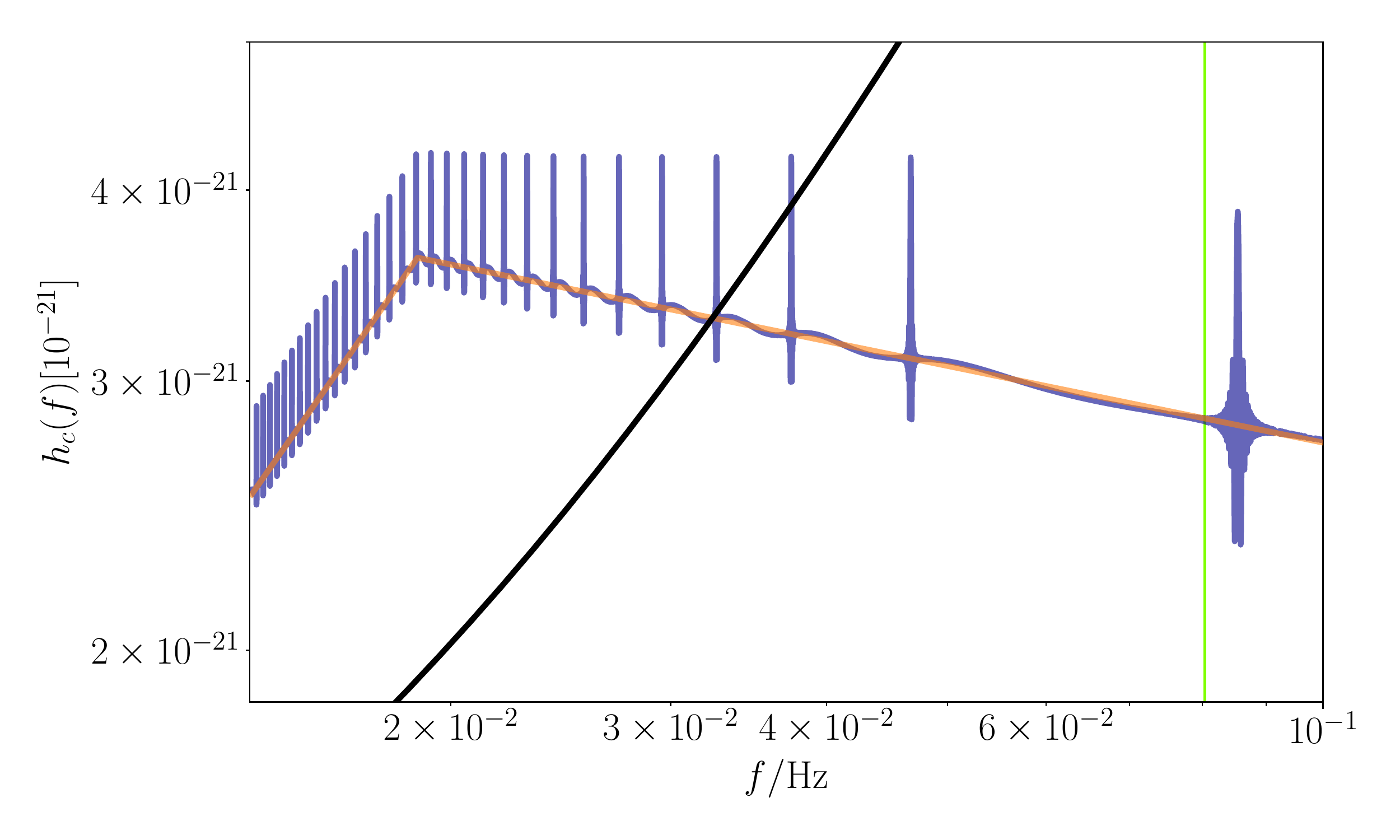} &
\includegraphics[scale=0.35]{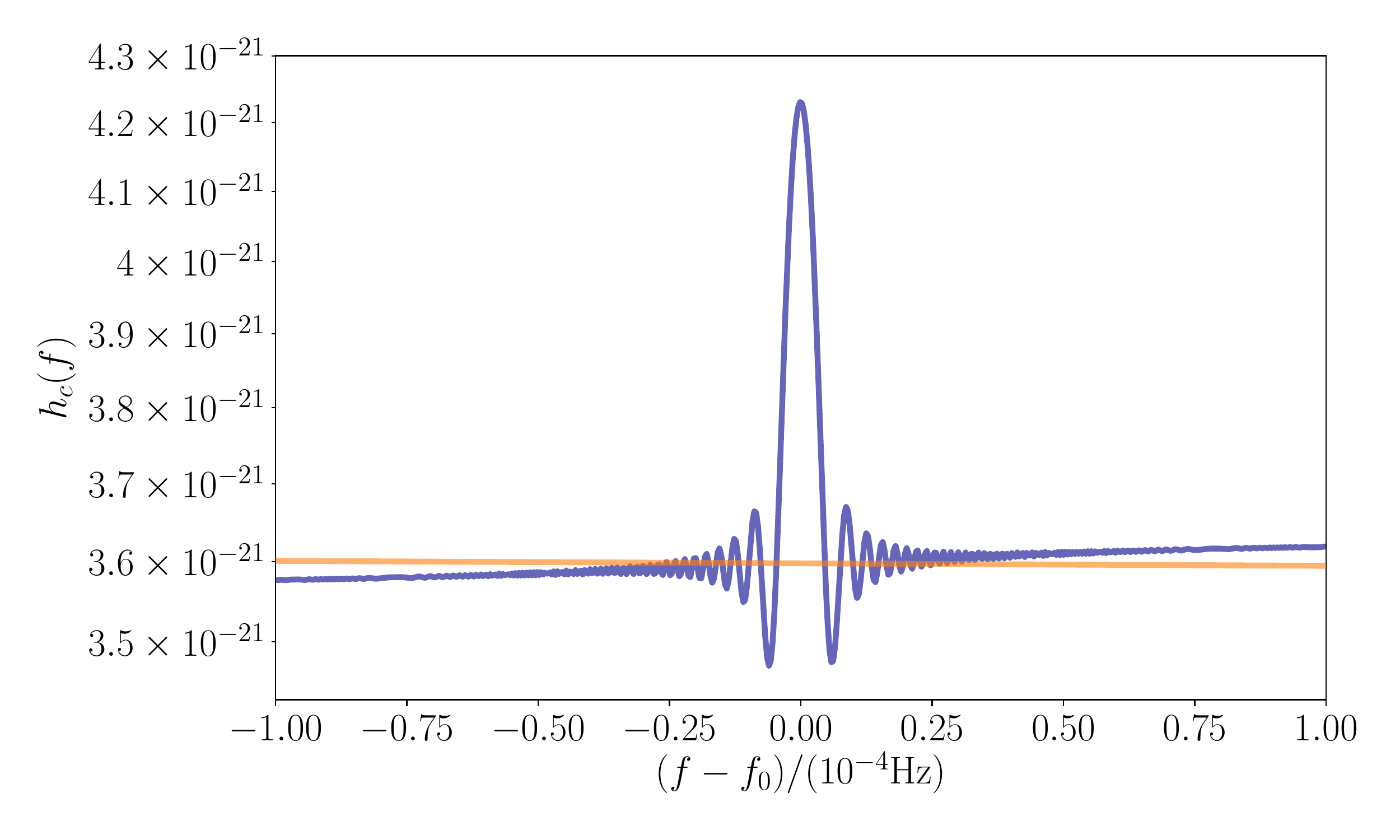} 
\end{array}$
\end{center}
\vspace{0pt}
\caption{
Successive zoom-in on the repeating-lens case of Figure \ref{Fig:hcflens}. The left panel captures all 32 lensing events and the right panel zooms in on the event with the highest SNR in LISA, which occurs at $\chi\approx0.2$, approaching the geometrical optics regime. The duration of the lensed signal in the right panel is approximately 14 hours.
}
\label{Fig:zoom}
\end{figure*}

In Figure \ref{Fig:hcflens}, we plot the characteristic strain vs. GW
frequency tracks of possible lensed BBH inspirals. The orange curves are the
unlensed sky and polarization averaged characteristic strains for BBHs on
circular orbits, with no proper motion with respect to the observer. That is, the orange tracks have characteristic strain,
\begin{equation}
h_{c,0} = \frac{8\pi^{2/3}}{\sqrt{10}} \frac{G^{5/3}}{c^4} \frac{\Mc^{5/3}_i f^{2/3}_{\GW,i} }{d(z)} \min\left[\sqrt{f_{\GW,i}\tau_{\text{LISA}}}, \sqrt{\frac{f^2_i}{\dot{f}_i}}\right],
\end{equation}
where $\Mc_i$ $f_{\GW,i}$, and $f_i$ are the redshifted quantities and so $d(z)$ is the
luminosity distance to the source. The quantities in brackets denote the
number of observed cycles per frequency bin given the LISA mission lifetime
$\tau_{\text{LISA}}$, which we assume to be $8$~years. Plotting this quantity
gives a good by-eye estimate of the signal to noise ratio \citep[see
\textit{e.g.}, Ref.][]{Sesana+2005, SD3:2018}. We plot each of the two orange curves
for a $M_i=50 \Msun$, $q_i=1$ binary, but at two different redshifts, $z=0.05$
and $z=0.2$, for visualization purposes.

The overplotted purple curves are the lensed and Doppler-boosted versions of
the orange curves in the stationary-lens
(upper curve) and repeating-lens (lower curve) regimes. These are computed as,
\begin{equation}
h_{c,\obs} = D  F(\chi) h_{c,0}
\end{equation}
where the time dependent Doppler factor, $D=\left[\gamma_o \left( 1 -
\beta_{||,o}(t) \right) \right]^{-1}$, for a Lorentz factor $\gamma_o$ and line
of sight orbital speed $\beta_{||,o}(t)$ of the outer binary, and $F(\chi)$ is
the lensing magnification given by Eq. (\ref{Eq:Mag1}). While the GW amplitude
is not affected to first order by a boost, this factor of $D$ appears in the
frequency and chirp mass terms just as the cosmological redshift does (see the
Appendix). We note that we have neglected the Doppler boost of the GW
frequency that appears in the lensing magnification (through $\chi$) because
lensing only occurs when the line of sight velocity of the orbit crosses zero.
For simplicity in describing the lensing signature, we also neglect $\ll1\%$
finite light-travel-time effects  due to the extent of the outer orbit.

The upper curve in Figure \ref{Fig:hcflens} ($z=0.05$, stationary lens),
assumes $M_L = 10^5\Msun$ in the stationary-lens regime with a
source position of $u=1$. While this case has a low probability for
occurrence, it is the GW-lensing scenario most commonly studied -- see similar
magnification curves in \citet{ChristianLoebVitale2018, TakaNaka:2003} -- and
we present it here to contrast with the repeating-lens scenario.  For
reference, we draw lines of constant $\chi=0.1, 1, 10$ in green, denoting the
wave-optics regime to the left, and the geometrical optics regime to the right
of the $\chi=1$ line. For lower GW frequencies, earlier in the BBH evolution,
the emitted GWs are longer than the size of the lens BH and are diffracted,
resulting in low magnification of the inner BBH GW emission. As the binary
frequency evolves to higher values, the full geometrical optics limit
magnification is reached and interference causes oscillation of the
lensing magnification at an increasing frequency given by Eqs. (\ref{Eq:Mag2}) and
(\ref{Eq:Magtd}). Interestingly, the wave-optics to geometrical optics
transition occurs just as the binary leaves the LISA band.

The lower purple curve ($z=0.2$, repeating lens) assumes a lens SMBH with mass
$M_L = 10^5\Msun$, and outer-orbital period of $P_o=0.25$~yr, squarely in the
repeating-lens regime for a multi-year LISA mission. The binary inclination is
set so that the  closest separation between source and lens is one Einstein
radius ($u=1$).  This case corresponds to a lensing probability of
$\mathcal{P}\sim0.01$, denoted by the purple `x' in Figure \ref{Fig:tscales}.
In this case, lensing manifests as repeated, symmetric spikes in the GW
amplitude that last for of order $0.5\mathcal{P}_{rl}P_o = 14$~hrs each, for
the example drawn here. Figure \ref{Fig:zoom} zooms in on these lensing
flares. In the left panel of Figure \ref{Fig:zoom}, one can see the much
smaller Doppler-boost oscillation between lensing flares (essentially a 
time-variable redshift), crossing $D=1$ during the lensing flare, as the line-of
sight orbital velocity crosses zero. The right panel of Figure \ref{Fig:zoom}
zooms in on the flare with the highest SNR, at the `knee' of the strain track
near $2 \times 10^{-2}$~Hz. Depending on when the final lensing event occurs
with respect to merger, such a flare could occur in the LIGO band, but with
probability $(\tau_{LIGO}/P_o) \mathcal{P}_{rl}$ instead of
$\mathcal{P}_{rl}$.

We note that in each of the lensing events drawn in Figure \ref{Fig:hcflens},
we have chosen the marginal source and lens separation for our definition of a
strong-lensing event, $u=1$, and a lens mass on the small end of the SMBH
spectrum. Because the lensing magnification increases with smaller $u$ and
larger lens mass (through $\chi$), these strain tracks are conservative
examples of possible lensing signatures.

\section{Astrophysical Population}
\label{S:Astrophysical Population}

We have shown that strong lensing of GWs from a BBH in a hierarchal triple
system can occur at a few to a few tens of percent of systems with outer
orbits that are comparable to or shorter than the detectable lifetime
($\tau_{\obs}$) of the inspiraling inner binary. We have also shown that
lensing causes a unique signature in the GW amplitude that can manifest from
the time-dependent magnification due to the moving and frequency-evolving GW
source. We now estimate how many such systems should exist based on possible
BBH populations.

\subsection{General Consideration}

We first take an approach agnostic to specific formation channels and assume
only that some fraction of BBH mergers occur in the clusters surrounding SMBHs
and that the number of BBHs per outer-orbital semi-major axis,
$dN_{\text{BBH}}/da_o$, that will merge, follows a power-law distribution in
the outer binary semi-major axis $\propto a^{\gamma}_o$, but depleted at
close separations by GW decay into the SMBH. That is,
\begin{equation}
\frac{dN_{\text{BBH}}}{da_o} \propto a^{\gamma}_o \min\left[ \frac{t_{\GW}(M_L, M_i, a_o)}{t_H}, 1 \right],
\end{equation}
where $t_H$ is the Hubble time, and we assume circular outer-binary orbits
(however, see \S\ref{S:Discussion}). Below we will consider a Bahcall-Wolf
cusp (with a stellar number density proportional to $a^{-7/4}_o$ implying
$\gamma=1/4$), \citep[Ref.][]{BW76, BW77} and also a distribution uniform in
log separation (number density proportional to $a^{-3}_o$ implying
$\gamma=-1$) motivated by the models of Ref. \cite{OlearyKocLoeb:2009}.

Given the BBH merger rate $\mathcal{R}_{\text{BBH}}$, probability
$\mathcal{P}$ for lensing as a function of distance from the SMBH $a_o$ and
SMBH mass $M_L$, and the fraction of mergers $\xi$ that occur in nuclear
clusters with the assumed BBH distribution, we compute the lensed merger rate
for a given lensing regime,
\begin{widetext}
\begin{equation}
\mathcal{R}_{\text{lens}, *} = \xi \mathcal{R}_{\text{BBH}} \frac{ \int_{M_{L}}\int^{a_{\max,*}}_{a_{\min,*}}{P(M_{L}) \frac{dN_{\text{BBH}}}{da_o}  \mathcal{P}(a_o, M_{L}) \ da_o \ dM_{L}} }{ \\\int_{M_{L}}\int^{r_{\text{infl}}}_{r_{\text{ISCO}}}{P(M_{L}) \frac{dN_{\text{BBH}}}{da_o} da_o \ dM_L}  } ,
\label{Eq:Rlens}
\end{equation}
\end{widetext}
where $a_{\min,*}$ to $a_{\max,*}$ is the range of outer-orbital separations
where the given lensing regime (denoted by *) is valid. The normalization
integrates over the entire distribution between SMBH ISCO and the radius of
influence of the BH, $r_{\text{infl}} = GM_L/\sigma^2$, for a stellar velocity dispersion
$\sigma$ that we take to be $100$~km/s (with our results being fairly insensitive
to this choice). The numerator also integrates over the SMBH population
by incorporating the BH mass function $P(M_{L})$, from \cite{Shankar:2004},
normalized to integrate to unity over the range $10^6-5 \times 10^{9} \Msun$.

We denote $\mathcal{R}_{\text{lens}, rl}$ the rate of occurrence of repeating-lens systems
where $a_{\min,rl}=r_{\text{ISCO}}(M_L)$ and $a_{\max,rl} = (\tau_{\obs}/(2
\pi))^{2/3} (GM_o)^{1/3}$. We denote $\mathcal{R}_{\text{lens}, s}$ the combined rate of
the slowly-moving and stationary lensing systems, where $a_{\min,s} = a_{\max,rl}$
and $a_{\max,s} = r_{\text{infl}}(M_L)$.

\begin{figure}
\begin{center}$
\begin{array}{c}
\includegraphics[scale=0.45]{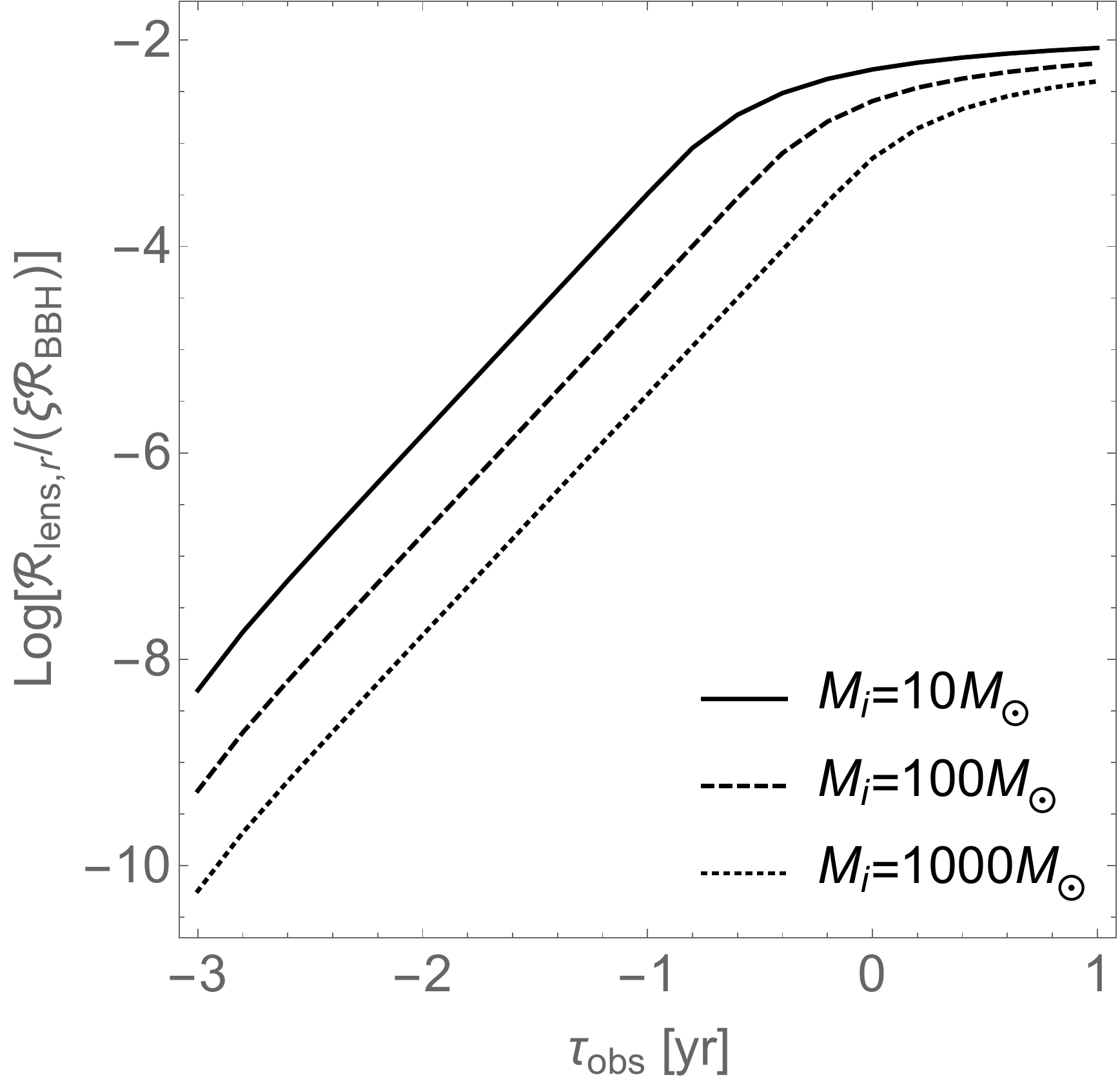} %&
\end{array}$
\end{center}
\vspace{0pt}
\caption{
Probability for repeating-lens events occurring around SHBs, assuming a constant-in-log-separation distribution of outer-binary orbits, and depleted at close separations due to GW decay of the outer binary. 
}
\label{Fig:Plens}
\end{figure}

Figure \ref{Fig:Plens} plots $\mathcal{R}_{\text{lens}, rl}$, in units of the
number of BBH mergers that occur in galactic nuclei $\xi
\mathcal{R}_{\text{BBH}}$, vs. the time $\tau_{\obs}$ that the inner BBH can
be observed in GWs, for the case of the $\gamma=-1$ merger distribution. The
lensing rate increases linearly with $\tau_{\obs}$, and then saturates for
observation times above approximately one year. This is because outer orbits
that are shorter than about $P_o \sim 0.13$~yr (for $M_L\geq 10^5 \Msun$ and
$M_i=50\Msun$) will decay into the SMBH in less than a Hubble time, and hence
decrease $dN_{\text{BBH}}/da_o$.

For fiducial values of $\sigma=100$~km/s, $q_i=1$, and $M_i=50\Msun$, and assuming a range of Bahcall-Wolf cusp to log-uniform number per outer binary separation distributions of BBHs, the limiting value of the lensing rate is, for $\tau_{\obs} \gtrsim 1$~yr,
\begin{eqnarray}
\begin{array}{l}
\dfrac{\mathcal{R}_{\text{lens}, rl} }{ \xi \mathcal{R}_{\text{BBH}}  }; \rightarrow 0.0001\%-0.7 \% \nonumber \\ \\ %\qquad  \\
\dfrac{\mathcal{R}_{\text{lens}, s} }{ \xi \mathcal{R}_{\text{BBH}}  }; \rightarrow 0.0002\%-0.007 \% 
\end{array}
\quad \gamma = \left\{1/4, -1\right\} .
\end{eqnarray}
Therefore for the steeper, $\gamma=-1$ distributions, lensing probabilities
reach the percent level, and are found in the rapidly
moving source regime, which has not been considered in previous works. Note
that while the Bahcall-Wolf cusp predicts many more possible BBHs far from
the SMBH, in the stationary-lens regime, the lensing probabilities are much
lower, and so the only high-lensing probability scenarios arise for
steep BBH distributions in the repeating lens regime.  This estimate implies
that for the $\gamma=-1$ case, if at least $150$ BBH systems with $\gtrsim
1$~yr to merge are found in LISA, at least one would exhibit a time-dependent
lensing flare in the GW amplitude.

%%%%%%%%%%%%%%%%%%%%%%%%%%%%%%%%%%%%%%%%%%%%%%%%
%%% FIGURE %%%
%%%%%%%%%%%%%%%%%%%%%%%%%%%%%%%%%%%%%%%%%%%%%%%%    

\subsection{Specific Channels}
\subsubsection{Single-Single Capture in Nuclear Star Clusters}
\label{S:APSS}

Ref. \cite{OlearyKocLoeb:2009} compute the distribution of mergers due to  BH-
BH GW induced captures in the BH cusps of nuclear star clusters. They find
that the distribution of mergers is roughly constant per log distance from the
BH. They also find $\sim 30\%$ of such BBH mergers occur within $10^4
GM_{L}/c^2$ of the central SMBH. This projection is consistent with our simple
approximation for $\gamma=-1$ in the previous section. Hence, one could
interpret the results of the previous subsection as predicting that one
percent of GW-capture mergers will be lensed in the LISA band.  An important
point to consider, however, is that the lifetime of BBHs formed in such
extreme dynamical channels will be greatly shortened due to the high initial
eccentricities required for single-single BH capture. If the BBH lifetime
falls below the period of the outer-binary orbit, then the lensing probability
also decreases.

The maximum lifetime of single-single BBH mergers is given in
\cite{OlearyKocLoeb:2009} as $t_{\text{mrg},i}\leq 4 \pi G  M_i / v^3_{\text{rel}}$,
where $v_{\text{rel}}$ is the relative velocity between two BHs before capture.
For merger times shorter than the outer-orbital period, we must decrease our
rate estimate above by a factor of $t_{\text{mrg},i}/P_o$. Taking the average
relative velocity between BHs in the nuclear star cluster to be on average the
circular velocity, $v_{\text{rel}} \sim \sqrt{G M_L/a_o} \sim 1400$~km/s, at
$P_o=1$~yr and $M_L=10^5 \Msun$, and using our fiducial BBH mass of $50
\Msun$, we see that single-single captures with such high relative velocity
will only last for $\sim9$ hours -- effectively eliminating chances to observe
lensing for these systems because such eccentric mergers will ensue before
completing an orbit around the central SMBH. They may also not appear in LISA
at all due to their high eccentricities \citep{OlearyKocLoeb:2009,
SamsingSS+2019, ChenAmaro:2017}. If, however, the relative velocity is
estimated at the velocity dispersion of $\sim100$~km/s, then the lifetime of
the inner binary is again of order years, predicting eccentric lensed
mergers in the LISA band. A combination of eccentricity and the appearance of
lensing signatures could serve as a probe of the properties of nuclear
clusters in the single-single GW capture scenario.

%%%%%%%%%%%%%%%%%%%%%%%%%%%%%%%%%%%%%%%%%%%%%%%%
%%% FIGURE %%%
%%%%%%%%%%%%%%%%%%%%%%%%%%%%%%%%%%%%%%%%%%%%%%%%    
\begin{figure}
\begin{center}%$
%\begin{array}{c}
\includegraphics[scale=0.27]{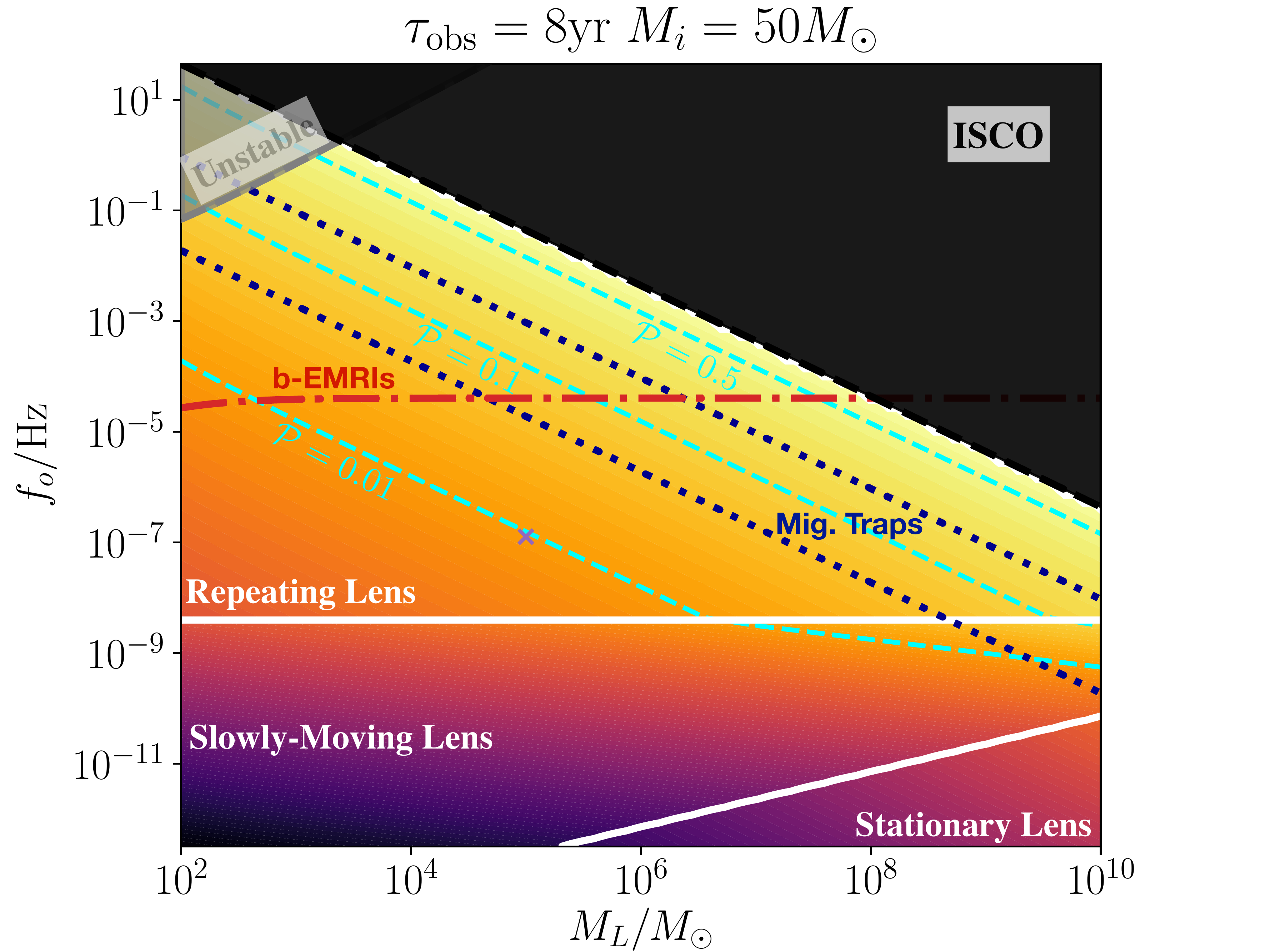} %&
%\end{array}$
\end{center}
\vspace{-10pt}
\caption{
The same as Figure \ref{Fig:tscales}, except we draw characteristic locations of the outer-binary orbit in the migration trap (dark-blue dashed lines) and binary-EMRI (red dot-dashed line) scenarios discussed in the text.
}
\label{Fig:tscales2}
\end{figure}

\subsubsection{The AGN channel and migration traps}

Within the AGN channel for merging BBHs
\citep[\textit{e.g.},][]{McKernanBBHAGN+2018}, gaseous torques can migrate BHs
through an AGN disk. Within some models of AGN disks, peaks in the disk
density profile can cause torques with opposite signs outward and inward of
the density peak. This causes a radial migration trap that can bring two BHs
to the location of the trap where they can merge, and even build up larger,
second generation mergers \citep{Bellovary+2016, Secunda+2018, YangBartos+2019}.  Figure
\ref{Fig:tscales2} replicates Figure \ref{Fig:tscales} but includes the
location of two possible migration traps from the literature, drawn as 
dark-blue dotted lines. These traps are located at $49$ and $662$ gravitational
radii from the central SMBH \citep{SirkoGoodman:2003}. If the migration traps
can hold the outer binary, and if gas forces do not merge the inner binary too
quickly\footnote{Generally, the effect of a gas on the binary orbit is still
an unsolved problem \citep[\textit{e.g.},][]{Tang+2017, MunozLai+2019, DuffellCBDq+2019}}, then such migration trap binaries would
have a high probability of exhibiting repeating-GW-lens flares, as well as a
strong Doppler boost signature. Hence, lensing could also be a powerful
discriminator for migration traps, and the AGN channel in general.

\subsubsection{binary-EMRI capture}

In the tidal capture scenario, a binary black hole in the vicinity of a SMBH
passes close to the tidal break-up radius of the binary on an eccentric orbit.
Orbital energy is taken from the outer binary and traded to the inner binary,
causing the inner binary to become eccentric. The outer binary circularizes to
the original outer-pericenter distance of tens of SMBH gravitational radii,
while the inner binary circularizes more slowly and can be emitting GWs
in the LISA band. \citep{ChenHan_bEMRI:2018, ChenLiCao:2019}. Depending on the detectable
lifetime of the inner binary in the LISA band, such tidal captures would have
a high probability of being lensed because of the tight orbit of the inner BBH
around the SMBH. In Figure \ref{Fig:tscales2} we plot this b-EMRI capture
radius (red dot-dashed line) for our fiducial $50\Msun$ BBH with equal mass
components.

We note that in both the b-EMRI and migration trap channels, the outer binary can be compact
enough for detection of GWs by LISA or a future Deci-hertz GW detector.
Lensing and the Doppler boost of GWs could be used to link the inner and outer
binary GW signals to each other. We leave further exploration of this
fascinating prospect to future work.

\section{Discussion}
\label{S:Discussion}

%%%%%%%%%%%%%%%%%%%
%%formation channels

While the current predictions for the number of LISA stellar-mass BBHs
detectable with pre-warning ability for the ground-based detectors number only
in the few to tens, the BBH inspirals that can be uncovered in LISA at
lower SNR from known LIGO mergers could number in the hundreds
\citep{SesanaMultibandGW:2016, KremerLISA+2019, Gerosa+2019}. Hence, the best case 
$\mathcal{O}(1\%)$ lensing predictions of the previous section are promising.
The detection of a repeating-lens event would confirm that some BBHs merge in
the environments of nuclear star clusters, close to the central SMBH. Depending
on the occurrence rate, lensing detection would constrain the fraction of BBH
mergers taking place in nuclear star cluster, $\xi$. Coupled with measurements
of binary parameters such as eccentricity, lensing detections could also
constrain the processes within a nuclear star cluster that lead to merger. That is, 
each of the specific merger scenarios discussed in \S\ref{S:Astrophysical
Population} suggest different outer-binary orbital distributions as well as
expected inner-and outer-binary eccentricities.

For example, single-single capture, binary-tidal capture, and the Kozai-Lidov
mechanism can generate highly eccentric BBH inspirals in nuclear star clusters
\citep{OlearyKocLoeb:2009, ChenHan_bEMRI:2018, Nishizawa+2017,
RandallZhongZhiEcc:2018, GiacomoKL+2019}. However, so can resonant three-body
processes in the BH cores of globular clusters
\citep[\textit{e.g.}][]{Samsing:2018, RodPNCMC+2018, SD3:2018, SD2:2018}.
Hence, it could be difficult to use binary parameter statistics alone to
disentangle these different formation scenarios. However, a lensing detection
would allow measurement of the central SMBH mass
\citep[\textit{e.g.}][]{TakaNaka:2003} and imply that the merger is occurring
next to a SMBH in the nuclear cluster at the center of a galaxy, rather than
in a globular cluster. From Doppler and lensing information we will also know
the size (up to the redshift) and shape of the outer orbit, allowing to vet
the efficiency of secular effects on the BBH inspiral, or compare to
predictions from gas migration and binary tidal capture.

Even for the non-detection of lensing events, $\mathcal{O}(100)$ LISA
detections of unlensed BBHs would begin to constrain the fraction of BBH
mergers occurring in nuclear star clusters, and the distribution of mergers in
such clusters.

%%%%%%%%%%%%%%%%%%%
%% easier to detect

As this work aims simply to point out the novelty and non-negligible
occurrence of repeating-lens events, we have not computed differential SNRs
for GW signals with and without lensing, and we have not computed the accuracy
of parameter recovery from lensed waveforms. These of course, will depend on
the SNR of the event, and so recovery of the lensing signature will depend on
the redshift of the source. However, Ref. \cite{WongBerti+2019} show that for
triple systems similar to the example repeating-lens system displayed in
Figure \ref{Fig:hcflens}, and generally those with year outer-orbital
timescales, have a detectable Doppler-boost signature (for non-face-on 
outer-orbital orientations) for which the orbital period can be recovered at $1\%$
precision (for an $\text{SNR}=10$ detection). With a detection of the Doppler
boost and such a precise determination of the outer-orbital period, the time
of a putative flare can be specified for recovery from the LISA data stream.
Furthermore, a $\sim 1.5$ increase in amplitude over the course of hours to
days, for an already detected BBH inspiral, is very likely discernible as it
would be equivalent to a $\sim30\%$ change in the inferred chirp mass for an
unlensed waveform, whereas the chirp mass relative error determination is
projected to be of order one part in a million \citep{SesanaMultibandGW:2016}.
Analysis of detection and parameter recovery for repeating-lens events should
be carried out in future work.

%%%%%%%%%%%%%%%%%%%
% Inclusion of eccentricity

For simplicity, we have considered only circular orbits for both binaries.
However, both inner- and outer-binary orbits could be eccentric. Here we
briefly discuss how eccentricity could affect our results.  For an eccentric
binary, the decay time will be shorter for the same outer-orbital period,
meaning that the green line in Figure \ref{Fig:tscales} will be moved
downwards and the lensing probabilities presented in Figure \ref{Fig:Plens}
will saturate at longer observational timescales. Typically, for average outer
binary eccentricities of $e\lesssim 0.5$, the decrease in orbital decay time
is less then a factor of three and our results are not changed significantly.
Eccentricity of the outer binary will also affect the shape of the lensing
flares plus Doppler signature in a predictable way that is presented for EM
lensing in \cite{HuSpikey+2019}.

Eccentricity of the inner binary will also cause the inner binary to merge
more quickly. As discussed in \S\ref{S:APSS}, highly eccentric BBHs
formed through GW capture could merge more quickly than an outer-orbital time,
greatly reducing the lensing probability. These mergers will likely have an
eccentricity signature in the LIGO 3rd generation or DECIGO bands
\citep{OlearyKocLoeb:2009, SamsingSS+2019}; evidence of high eccentricity,
in conjunction with non-detection of lensing events, could corroborate such a
scenario.

Eccentric inner binaries will also emit GWs at multiple higher harmonics of
the binary orbital frequency. Hence, the same binary will be lensed
differently for different harmonics because the value of $\chi$ in the lensing
magnification (Eq. \ref{Eq:Mag1}) is different for each harmonic. For high-SNR
lensing events where multiple harmonics can be detected, propagation of GWs in
the wave and geometrical optics limit could be observed simultaneously.
We note, however, that the spread of GW power into higher harmonics
could result in lower SNRs for highly eccentric systems.

Finally, eccentricity of either binary will act to destabilize the triple,
resulting in the gray forbidden region in Figures \ref{Fig:tscales} and
\ref{Fig:tscales2} moving downwards. However, as stability only affects very
compact outer orbits, the shorter decay times of eccentric binaries will
likely affect our results before stability becomes an issue.

Further study of the effects of eccentricity and the eccentricity inducing
Kozai-Lidov mechanism on our results is warranted in a future study. It may be
especially important to include higher order relativistic effects
\citep[\textit{e.g.}][]{BinLiuGRKL+2019} in the regimes important for
repeated lensing.

Figure \ref{Fig:Plens} also shows the expected lensing rate for more massive
BBH mergers, up to $M_i=1000 \Msun$. While lensing of these more massive
mergers is less probable, it may be that many more can be detected by LISA.
LISA is the most sensitive to such $10^3-10^4 \Msun$ mergers and could detect
them out to redshifts of $z\sim10^2$ \citep{JaniCutler:2019}. If intermediate
mass BH (IMBH) mergers occur close to SMBHs in hierarchical triples,
\textit{e.g.} through IMBH delivery via devoured globular clusters
\citep{Ebisuzaki+2001, FragioneIMBHa:2018}, or through BH build-up in
migration traps \citep{Bellovary+2016}, then lensing would be a common feature
in the IMBH mergers detected in the LISA band.

At the opposite end of the mass spectrum, binary neutron star, or binary white
dwarf mergers could be detected by LISA within the galaxy
\citep{KremerMWLISA+2018}. Hence, if such mergers occur in nearby globular
clusters harboring a putative IMBH, or the galactic center, they could also be
lensed.

%%%%%%%%%%%%%%%%%%%
% test gravity

While we have stressed the utility of lensing events to elucidate
astrophysical channels for BBH formation and evolution, we note also that
observing repeating lensing of GWs emitted from a BBH as it evolves in orbital
frequency and possibly eccentricity could allow novel tests of GW propagation.

For example, the lens magnification was derived under the assumption that the
GWs propagate following a linear wave equation in the weak field. If this is
violated by, \textit{e.g.} MONDIAN theories of gravity that predict non-linear
propagation, then the magnification predicted here could also be altered
\citep[see \textit{e.g.}][]{CheslerLoeb:2017, NishizawaGWprop:2018}. Such
changes to the lensing waveforms can be worked out within a chosen modified
theory. Additionally, the oscillation of the lensed amplitude of GWs seen in
both types of lensing events in Figure \ref{Fig:hcflens} is set by the GW time
delay. If GWs do not travel at the speed of light, or interact with themselves
or the background spacetime non-linearly, one might expect this delay time to
be altered, and so directly observable from a GW lensing event.

Such tests may be especially interesting when lensing can be observed in both
the wave and geometrical optics regimes for the same GW source. Comparison of
the two regimes can occur at different times over the evolution of the BBH in
frequency, or in a simultaneous manner for eccentric binaries emitting GWs at
multiple harmonics.

We have not treated changes in the polarization tensor of the GW due to
lensing. These changes come in at the order of the lensing potential
experienced by the propagating GWs, which is assumed to be much smaller than
unity in the derivation of the lensing magnification \citep{TakaNaka:2003}. In
situations where the source of GWs is within tens of gravitational radii from
the SMBH however, polarization effects could enter at the few to ten percent
level and could offer another probe of the central SMBH spacetime.

%%%%%%%%%%%%%%%%%%%
% EM oppurtunities

Finally, we note that electromagnetic lensing could occur if the BBH emits
light, as might occur in the AGN channel. In that case, it is usually assumed
that the AGN outshines the accretion powered luminosity generated by the inner
BBH. However, if the lensing+Doppler period is already known, then a targeted
search for that AGN periodicity in the localized error box could help to
identify the weak periodicity and hence the host AGN. This would work
similarly to a lock-in amplifier where the signal dithering is generated by
the outer-binary orbit. The fractional amplitude of modulation that could be
recovered would be proportional to the precision at which the outer-binary
period can be determined from the GW lensing plus Doppler signature. This
would also allow a direct comparison of the lensing of light and GWs as a test
of the equivalence principle \citep{HaimanEMChirp:2017} and also isolate the
electromagnetic emission coming from accretion onto the BBH as opposed to
accretion onto the SMBH.

\section{Conclusions}
\label{S:Conclusion}

We have shown that inspiraling BBHs that emit GWs in the LISA band and orbit a
tertiary SMBH in a hierarchical triple with of order one year outer orbital
periods, have up to a percent probability of being repeatedly strongly lensed
by the SMBH. This lensing probability is only large when GWs from the inspiraling BBH
can be observed for the orbital period of the BBH around the SMBH or longer
and when the decay time of the BBH into the SMBH is of order a Hubble time or
longer. The balance of the two sets the required $\mathcal{O}(\text{yr})$
orbital timescales.  These short outer-orbital timescales can be populated
for steep distributions of BBHs around the central SMBH ($dN/da_0 \propto
a^{-1}_o$ or steeper) that have been shown to occur for single-single GW
capture scenarios or via migration traps in AGN disks, or through binary
tidal capture by the central SMBH. LISA has the best chance of discovering
such a repeating-lens event, because it will be able to detect inspiraling BBHs
for the final years of their lives and because it could detect up to hundreds of
such events \citep[][though fewer at high SNR]{SesanaMultibandGW:2016,
KremerMWLISA+2018, Gerosa+2019}.

The characteristic signature of these lensing events is magnification by a
factor of $\sim1.1\times$ to $2\times$ (cf. Eq. \ref{Eq:Mag1}) of the GW
amplitude for a duration of $\sim 0.01$ of the period of the BBH in its orbit
around the SMBH (hours to days). The event will repeat once every outer-binary
orbital period and be accompanied by detectable deviations in the GW phase due
to the relativistic Doppler boost \citep[\textit{e.g.},][]{WongBerti+2019}. A
detection of this phenomenon will provide the mass of the central SMBH,
helping to constrain BBH formation scenarios, specifically pointing to
formation in nuclear star clusters. Detection of time and frequency variable
lensing of GWs could also provide a new testing ground for theories of
gravity. Non-detection will put upper limits on the fraction of BBHs formed in
the vicinity of SMBHs and the astrophysical processes that form them.

\acknowledgements
We thank Bence Kocsis for insightful comments that improved the presentation
of this work. Financial support was provided from NASA through Einstein
Postdoctoral Fellowship award number PF6-170151 and funding from the Institute
for Theory and Computation Fellowship (DJD) and through the Black Hole
Initiative which is funded by grants from the John Templeton Foundation and
the Gordon and Betty Moore Foundation.

\appendix
\section{Effect of the Doppler Boost on the GW Amplitude }

We sketch a derivation of the relativistic Doppler boost formula for GWs,
ignoring at first any cosmological redshift. We consider a GW in the time
domain with frequency $f$, amplitude $|h(t(f))|$ and phase $\phi(t(f))$ (for
notational simplicity, we will write $|h(f)|$ hereafter). We do not consider
the detection of different GW polarizations, although see Ref. 
\cite{Torres-Orjuela+2018}.

In the rest frame of the GW emitting source the GW frequency at any time is
given by $f_{\em}$. In the frame of an observer for which the source has a
relative speed $\beta = v/c$ with a fraction $\beta_{||}$ along the line of
sight, the observed GW frequency is $f_{\obs}=Df_{\em}$, where the Doppler
factor $D$ is,
\begin{align}
D &\equiv  \left[\gamma \left( 1 - \beta_{||}\right) \right]^{-1} 
\label{Eq:Ddef}
\end{align}
with $\gamma = \left(1-\beta^2\right)^{-1/2}$. Then the measured phase of the wave changes over time due to the Doppler boost. The phase accumulated over an observation time of length $\tau$ is
\begin{equation}
\phi(\tau) = \int^{t_0+\tau}_{t_0}{\pi Df_{\em} dt}. 
\label{Eq:DopPh}
\end{equation}
which has been considered in many other works \citep[see Ref.][and references
therein]{WongBerti+2019}

We also consider the effect of the Doppler boost on the wave amplitude. Just
as for electromagnetic waves, we use that the photon/graviton occupation
number is Lorentz invariant,
\begin{equation}
\frac{I_{E}}{E^3} =\frac{I_{f}}{f^3} = \text{Invar.} , %\qquad \text{(EM)}.
\label{EQ:SRInvar}
\end{equation}
where $E$ is the energy of a photon or graviton, $f$ is the wave frequency and
the first equality makes the assumption that the energy and frequency of a
gravitational wave are related linearly (if they are not, the
observed Doppler boost of GWs would recover this). $I_f$ is the specific
intensity of the radiation.

For GWs, occupation number is still a Lorentz invariant and the transformation
of the specific intensity still holds, but we are interested in the
transformation of the wave amplitude $|h(f)|$. The specific intensity and
$|h(f)|$ are related by
\begin{equation}
|h(f)|^2 = \frac{\pi c^3}{4 G} f^{-2}\frac{dE}{dA dt} = \frac{\pi c^3}{4 G f^2}\int{I_{f} \ d\Omega df},
\label{Eq:hdef}
\end{equation}
and for gravitational waves we can write,
\begin{equation}
I_f \equiv \frac{dE_n}{dAdtd\Omega df} = \frac{1}{4 \pi r^2(z)}\frac{dP}{ d\Omega} \delta\left( f -n f_K \right)
\end{equation}
where $dP/d\Omega$ is the angle dependent emitted power, $f_K$ is the
Keplerian orbital frequency, and GWs are emitted with frequency at the $n^{\text{th}}$ harmonic of $f_K$. Using that $I_{f \obs } = D^3 I_{f,\em}$,
$d\Omega_{\obs} = D^{-2}d\Omega_{\em}$, $df_{\obs} = D df_{\em}$ and
integrating we find,
\begin{eqnarray}
  |h(f_{\obs})|^2_{n,\obs} &=& |h (f_{\obs}/D)|^2_{n,\em}. 
  \label{Eq:hTrans}
\end{eqnarray}
This may also be found by simply transforming the first quantity in Eq.
(\ref{Eq:hdef}), using that $dE_{\obs} = D dE_{\em}$, $dA_{\obs}=dA_{\em}$,
and $dt_{\obs}= D^{-1} dt_{\em}$.

Dropping the $n$ subscript, the characteristic strain, $h^2_c(f) = |h(f)|^2
f^2/\dot{f}$, transforms the same as $|h(f)|$ because $\dot{f}_{\obs} = D^2
\dot{f}_{\em}$, that is, the number of cycles per frequency bin is invariant.
The Fourier transform of the strain amplitude is given by
$|\tilde{h}(f_{\obs})|^2_{\obs} = D^{-2}|\tilde{h}(f_{\obs}/D)|^2_{\em}$,
where the extra factor of $D^{-2}$ from Eq. (\ref{Eq:hTrans}) simply accounts
for the extra factor of $\dot{f}$ in $|\tilde{h}(f)|^2$.

Hence, the transformation of the strain amplitude does not explicitly contain
the Doppler factor, but the amplitude is altered by its dependence on the
frequency. This frequency dependence arises for the same reason that the
electromagnetic Doppler formula is written with a $D^{3-\alpha}$ dependence
for a specific intensity that goes as $f^{-\alpha}$, and for the same reason
that the redshifted GW amplitude depends on the luminosity distance and not
the co-moving distance.

We can then write the strain amplitude for the inner-binary in terms of
observed quantities as,
\begin{equation}
|h_i| \propto D \frac{\Mc^{5/3}_i f^{2/3}_i }{d(z)} g(I_i),
\label{Eq:hs}
\end{equation}
where $g(I_i)$ is the inclination dependence of the inner binary and $d(z)$ is
the luminosity distance to the source. This follows from the relations
\begin{align}
&f = \frac{D}{1+z}f_{\em}, \quad 
\dot{f} = \left(\frac{D}{1+z} \right)^2\dot{f}_{\em}, \nonumber \\
&\Mc \propto \dot{f}^{3/5} f^{-11/5}.
\label{Eq:fobs}
\end{align}

\bibliography{refs}

\begin{thebibliography}{67}
\expandafter\ifx\csname natexlab\endcsname\relax\def\natexlab#1{#1}\fi
\expandafter\ifx\csname bibnamefont\endcsname\relax
  \def\bibnamefont#1{#1}\fi
\expandafter\ifx\csname bibfnamefont\endcsname\relax
  \def\bibfnamefont#1{#1}\fi
\expandafter\ifx\csname citenamefont\endcsname\relax
  \def\citenamefont#1{#1}\fi
\expandafter\ifx\csname url\endcsname\relax
  \def\url#1{\texttt{#1}}\fi
\expandafter\ifx\csname urlprefix\endcsname\relax\def\urlprefix{URL }\fi
\providecommand{\bibinfo}[2]{#2}
\providecommand{\eprint}[2][]{\url{#2}}

\bibitem[{\citenamefont{{The LIGO Scientific Collaboration}
  et~al.}(2018)\citenamefont{{The LIGO Scientific Collaboration}, {the Virgo
  Collaboration}, {Abbott}, {Abbott}, {Abbott}, {Abraham}, {Acernese},
  {Ackley}, {Adams}, and {Adhikari}}}]{LIGO_O2_COBs:2018}
\bibinfo{author}{\bibnamefont{{The LIGO Scientific Collaboration}}},
  \bibinfo{author}{\bibnamefont{{the Virgo Collaboration}}},
  \bibinfo{author}{\bibfnamefont{B.~P.} \bibnamefont{{Abbott}}},
  \bibinfo{author}{\bibfnamefont{R.}~\bibnamefont{{Abbott}}},
  \bibinfo{author}{\bibfnamefont{T.~D.} \bibnamefont{{Abbott}}},
  \bibinfo{author}{\bibfnamefont{S.}~\bibnamefont{{Abraham}}},
  \bibinfo{author}{\bibfnamefont{F.}~\bibnamefont{{Acernese}}},
  \bibinfo{author}{\bibfnamefont{K.}~\bibnamefont{{Ackley}}},
  \bibinfo{author}{\bibfnamefont{C.}~\bibnamefont{{Adams}}}, \bibnamefont{and}
  \bibinfo{author}{\bibfnamefont{R.~X.} \bibnamefont{{Adhikari}}},
  \bibinfo{journal}{arXiv e-prints} \bibinfo{eid}{arXiv:1811.12907}
  (\bibinfo{year}{2018}), \eprint{1811.12907}.

\bibitem[{\citenamefont{{The LIGO Scientific Collaboration} and {the Virgo
  Collaboration}}(2018)}]{LIGO_O2_BBHproperties:2018}
\bibinfo{author}{\bibnamefont{{The LIGO Scientific Collaboration}}}
  \bibnamefont{and} \bibinfo{author}{\bibnamefont{{the Virgo Collaboration}}},
  \bibinfo{journal}{arXiv e-prints} \bibinfo{eid}{arXiv:1811.12940}
  (\bibinfo{year}{2018}), \eprint{1811.12940}.

\bibitem[{\citenamefont{{Yunes} et~al.}(2011)\citenamefont{{Yunes}, {Kocsis},
  {Loeb}, and {Haiman}}}]{YunesKocsisLoeb:2011}
\bibinfo{author}{\bibfnamefont{N.}~\bibnamefont{{Yunes}}},
  \bibinfo{author}{\bibfnamefont{B.}~\bibnamefont{{Kocsis}}},
  \bibinfo{author}{\bibfnamefont{A.}~\bibnamefont{{Loeb}}}, \bibnamefont{and}
  \bibinfo{author}{\bibfnamefont{Z.}~\bibnamefont{{Haiman}}},
  \bibinfo{journal}{\prl} \textbf{\bibinfo{volume}{107}}, \bibinfo{eid}{171103}
  (\bibinfo{year}{2011}), \eprint{1103.4609}.

\bibitem[{\citenamefont{{Kocsis} et~al.}(2011)\citenamefont{{Kocsis}, {Yunes},
  and {Loeb}}}]{KocsisYunesLoeb:2011}
\bibinfo{author}{\bibfnamefont{B.}~\bibnamefont{{Kocsis}}},
  \bibinfo{author}{\bibfnamefont{N.}~\bibnamefont{{Yunes}}}, \bibnamefont{and}
  \bibinfo{author}{\bibfnamefont{A.}~\bibnamefont{{Loeb}}},
  \bibinfo{journal}{\prd} \textbf{\bibinfo{volume}{84}}, \bibinfo{eid}{024032}
  (\bibinfo{year}{2011}), \eprint{1104.2322}.

\bibitem[{\citenamefont{{Fedrow} et~al.}(2017)\citenamefont{{Fedrow}, {Ott},
  {Sperhake}, {Blackman}, {Haas}, {Reisswig}, and {De Felice}}}]{Fedrow+2017}
\bibinfo{author}{\bibfnamefont{J.~M.} \bibnamefont{{Fedrow}}},
  \bibinfo{author}{\bibfnamefont{C.~D.} \bibnamefont{{Ott}}},
  \bibinfo{author}{\bibfnamefont{U.}~\bibnamefont{{Sperhake}}},
  \bibinfo{author}{\bibfnamefont{J.}~\bibnamefont{{Blackman}}},
  \bibinfo{author}{\bibfnamefont{R.}~\bibnamefont{{Haas}}},
  \bibinfo{author}{\bibfnamefont{C.}~\bibnamefont{{Reisswig}}},
  \bibnamefont{and} \bibinfo{author}{\bibfnamefont{A.}~\bibnamefont{{De
  Felice}}}, \bibinfo{journal}{\prl} \textbf{\bibinfo{volume}{119}},
  \bibinfo{eid}{171103} (\bibinfo{year}{2017}), \eprint{1704.07383}.

\bibitem[{\citenamefont{{D'Orazio} and {Loeb}}(2018)}]{DOrazioLoebSP:2018}
\bibinfo{author}{\bibfnamefont{D.~J.} \bibnamefont{{D'Orazio}}}
  \bibnamefont{and} \bibinfo{author}{\bibfnamefont{A.}~\bibnamefont{{Loeb}}},
  \bibinfo{journal}{\prd} \textbf{\bibinfo{volume}{97}}, \bibinfo{eid}{083008}
  (\bibinfo{year}{2018}), \eprint{1706.04211}.

\bibitem[{\citenamefont{{Derdzinski} et~al.}(2019)\citenamefont{{Derdzinski},
  {D'Orazio}, {Duffell}, {Haiman}, and {MacFadyen}}}]{Derdzinski+2018}
\bibinfo{author}{\bibfnamefont{A.~M.} \bibnamefont{{Derdzinski}}},
  \bibinfo{author}{\bibfnamefont{D.}~\bibnamefont{{D'Orazio}}},
  \bibinfo{author}{\bibfnamefont{P.}~\bibnamefont{{Duffell}}},
  \bibinfo{author}{\bibfnamefont{Z.}~\bibnamefont{{Haiman}}}, \bibnamefont{and}
  \bibinfo{author}{\bibfnamefont{A.}~\bibnamefont{{MacFadyen}}},
  \bibinfo{journal}{\mnras} \textbf{\bibinfo{volume}{486}},
  \bibinfo{pages}{2754} (\bibinfo{year}{2019}), \eprint{1810.03623}.

\bibitem[{\citenamefont{{LIGO Scientific Collaboration} and {the Virgo
  Collaboration}}(2018)}]{LIGOBNSEOS:2018}
\bibinfo{author}{\bibnamefont{{LIGO Scientific Collaboration}}}
  \bibnamefont{and} \bibinfo{author}{\bibnamefont{{the Virgo Collaboration}}},
  \bibinfo{journal}{\prl} \textbf{\bibinfo{volume}{121}}, \bibinfo{eid}{161101}
  (\bibinfo{year}{2018}), \eprint{1805.11581}.

\bibitem[{\citenamefont{{Meiron} et~al.}(2017)\citenamefont{{Meiron}, {Kocsis},
  and {Loeb}}}]{MeironBenceLoeb:2017}
\bibinfo{author}{\bibfnamefont{Y.}~\bibnamefont{{Meiron}}},
  \bibinfo{author}{\bibfnamefont{B.}~\bibnamefont{{Kocsis}}}, \bibnamefont{and}
  \bibinfo{author}{\bibfnamefont{A.}~\bibnamefont{{Loeb}}},
  \bibinfo{journal}{\apj} \textbf{\bibinfo{volume}{834}}, \bibinfo{eid}{200}
  (\bibinfo{year}{2017}), \eprint{1604.02148}.

\bibitem[{\citenamefont{{Kocsis} et~al.}(2006)\citenamefont{{Kocsis}, {Frei},
  {Haiman}, and {Menou}}}]{KocsisStndSiren:2006}
\bibinfo{author}{\bibfnamefont{B.}~\bibnamefont{{Kocsis}}},
  \bibinfo{author}{\bibfnamefont{Z.}~\bibnamefont{{Frei}}},
  \bibinfo{author}{\bibfnamefont{Z.}~\bibnamefont{{Haiman}}}, \bibnamefont{and}
  \bibinfo{author}{\bibfnamefont{K.}~\bibnamefont{{Menou}}},
  \bibinfo{journal}{\apj} \textbf{\bibinfo{volume}{637}}, \bibinfo{pages}{27}
  (\bibinfo{year}{2006}), \eprint{astro-ph/0505394}.

\bibitem[{\citenamefont{{Bonvin} et~al.}(2017)\citenamefont{{Bonvin},
  {Caprini}, {Sturani}, and {Tamanini}}}]{Bonvin+2017}
\bibinfo{author}{\bibfnamefont{C.}~\bibnamefont{{Bonvin}}},
  \bibinfo{author}{\bibfnamefont{C.}~\bibnamefont{{Caprini}}},
  \bibinfo{author}{\bibfnamefont{R.}~\bibnamefont{{Sturani}}},
  \bibnamefont{and}
  \bibinfo{author}{\bibfnamefont{N.}~\bibnamefont{{Tamanini}}},
  \bibinfo{journal}{\prd} \textbf{\bibinfo{volume}{95}}, \bibinfo{eid}{044029}
  (\bibinfo{year}{2017}), \eprint{1609.08093}.

\bibitem[{\citenamefont{{Chen} et~al.}(2019)\citenamefont{{Chen}, {Li}, and
  {Cao}}}]{ChenLiCao:2019}
\bibinfo{author}{\bibfnamefont{X.}~\bibnamefont{{Chen}}},
  \bibinfo{author}{\bibfnamefont{S.}~\bibnamefont{{Li}}}, \bibnamefont{and}
  \bibinfo{author}{\bibfnamefont{Z.}~\bibnamefont{{Cao}}},
  \bibinfo{journal}{\mnras} \textbf{\bibinfo{volume}{485}},
  \bibinfo{pages}{L141} (\bibinfo{year}{2019}), \eprint{1703.10543}.

\bibitem[{\citenamefont{{Inayoshi} et~al.}(2017)\citenamefont{{Inayoshi},
  {Tamanini}, {Caprini}, and {Haiman}}}]{Inayoshi+Hamian:2017}
\bibinfo{author}{\bibfnamefont{K.}~\bibnamefont{{Inayoshi}}},
  \bibinfo{author}{\bibfnamefont{N.}~\bibnamefont{{Tamanini}}},
  \bibinfo{author}{\bibfnamefont{C.}~\bibnamefont{{Caprini}}},
  \bibnamefont{and} \bibinfo{author}{\bibfnamefont{Z.}~\bibnamefont{{Haiman}}},
  \bibinfo{journal}{\prd} \textbf{\bibinfo{volume}{96}}, \bibinfo{eid}{063014}
  (\bibinfo{year}{2017}), \eprint{1702.06529}.

\bibitem[{\citenamefont{{Robson} et~al.}(2018)\citenamefont{{Robson},
  {Cornish}, {Tamanini}, and {Toonen}}}]{RobsonTriples+2018}
\bibinfo{author}{\bibfnamefont{T.}~\bibnamefont{{Robson}}},
  \bibinfo{author}{\bibfnamefont{N.~J.} \bibnamefont{{Cornish}}},
  \bibinfo{author}{\bibfnamefont{N.}~\bibnamefont{{Tamanini}}},
  \bibnamefont{and} \bibinfo{author}{\bibfnamefont{S.}~\bibnamefont{{Toonen}}},
  \bibinfo{journal}{\prd} \textbf{\bibinfo{volume}{98}}, \bibinfo{eid}{064012}
  (\bibinfo{year}{2018}), \eprint{1806.00500}.

\bibitem[{\citenamefont{{Wong} et~al.}(2019)\citenamefont{{Wong}, {Baibhav},
  and {Berti}}}]{WongBerti+2019}
\bibinfo{author}{\bibfnamefont{K.~W.~K.} \bibnamefont{{Wong}}},
  \bibinfo{author}{\bibfnamefont{V.}~\bibnamefont{{Baibhav}}},
  \bibnamefont{and} \bibinfo{author}{\bibfnamefont{E.}~\bibnamefont{{Berti}}},
  \bibinfo{journal}{\mnras} \textbf{\bibinfo{volume}{488}},
  \bibinfo{pages}{5665} (\bibinfo{year}{2019}), \eprint{1902.01402}.

\bibitem[{\citenamefont{{Randall} and {Xianyu}}(2019)}]{RandallXianyu:2019}
\bibinfo{author}{\bibfnamefont{L.}~\bibnamefont{{Randall}}} \bibnamefont{and}
  \bibinfo{author}{\bibfnamefont{Z.-Z.} \bibnamefont{{Xianyu}}},
  \bibinfo{journal}{\apj} \textbf{\bibinfo{volume}{878}}, \bibinfo{eid}{75}
  (\bibinfo{year}{2019}), \eprint{1805.05335}.

\bibitem[{\citenamefont{{Takahashi} and {Nakamura}}(2003)}]{TakaNaka:2003}
\bibinfo{author}{\bibfnamefont{R.}~\bibnamefont{{Takahashi}}} \bibnamefont{and}
  \bibinfo{author}{\bibfnamefont{T.}~\bibnamefont{{Nakamura}}},
  \bibinfo{journal}{\apj} \textbf{\bibinfo{volume}{595}}, \bibinfo{pages}{1039}
  (\bibinfo{year}{2003}), \eprint{astro-ph/0305055}.

\bibitem[{\citenamefont{{Ruffa}}(1999)}]{Ruffa:1999}
\bibinfo{author}{\bibfnamefont{A.~A.} \bibnamefont{{Ruffa}}},
  \bibinfo{journal}{\apjl} \textbf{\bibinfo{volume}{517}}, \bibinfo{pages}{L31}
  (\bibinfo{year}{1999}).

\bibitem[{\citenamefont{{Christian} et~al.}(2018)\citenamefont{{Christian},
  {Vitale}, and {Loeb}}}]{ChristianLoebVitale2018}
\bibinfo{author}{\bibfnamefont{P.}~\bibnamefont{{Christian}}},
  \bibinfo{author}{\bibfnamefont{S.}~\bibnamefont{{Vitale}}}, \bibnamefont{and}
  \bibinfo{author}{\bibfnamefont{A.}~\bibnamefont{{Loeb}}},
  \bibinfo{journal}{\prd} \textbf{\bibinfo{volume}{98}}, \bibinfo{eid}{103022}
  (\bibinfo{year}{2018}), \eprint{1802.02586}.

\bibitem[{\citenamefont{{Kocsis}}(2013)}]{KocsisLens:2013}
\bibinfo{author}{\bibfnamefont{B.}~\bibnamefont{{Kocsis}}},
  \bibinfo{journal}{\apj} \textbf{\bibinfo{volume}{763}}, \bibinfo{eid}{122}
  (\bibinfo{year}{2013}), \eprint{1211.6427}.

\bibitem[{\citenamefont{{Antonini} and {Perets}}(2012)}]{AntoniniPerets:2012}
\bibinfo{author}{\bibfnamefont{F.}~\bibnamefont{{Antonini}}} \bibnamefont{and}
  \bibinfo{author}{\bibfnamefont{H.~B.} \bibnamefont{{Perets}}},
  \bibinfo{journal}{\apj} \textbf{\bibinfo{volume}{757}}, \bibinfo{eid}{27}
  (\bibinfo{year}{2012}), \eprint{1203.2938}.

\bibitem[{\citenamefont{{Chen} and {Han}}(2018)}]{ChenHan_bEMRI:2018}
\bibinfo{author}{\bibfnamefont{X.}~\bibnamefont{{Chen}}} \bibnamefont{and}
  \bibinfo{author}{\bibfnamefont{W.-B.} \bibnamefont{{Han}}},
  \bibinfo{journal}{Condensed Matter Physics} \textbf{\bibinfo{volume}{1}},
  \bibinfo{eid}{53} (\bibinfo{year}{2018}), \eprint{1801.05780}.

\bibitem[{\citenamefont{{Fern{\'a}ndez} and {Kobayashi}}(2018)}]{FernKoby:2018}
\bibinfo{author}{\bibfnamefont{J.~J.} \bibnamefont{{Fern{\'a}ndez}}}
  \bibnamefont{and}
  \bibinfo{author}{\bibfnamefont{S.}~\bibnamefont{{Kobayashi}}},
  \bibinfo{journal}{arXiv e-prints}  (\bibinfo{year}{2018}),
  \eprint{1805.09593}.

\bibitem[{\citenamefont{{Bellovary} et~al.}(2016)\citenamefont{{Bellovary},
  {Mac Low}, {McKernan}, and {Ford}}}]{Bellovary+2016}
\bibinfo{author}{\bibfnamefont{J.~M.} \bibnamefont{{Bellovary}}},
  \bibinfo{author}{\bibfnamefont{M.-M.} \bibnamefont{{Mac Low}}},
  \bibinfo{author}{\bibfnamefont{B.}~\bibnamefont{{McKernan}}},
  \bibnamefont{and} \bibinfo{author}{\bibfnamefont{K.~E.~S.}
  \bibnamefont{{Ford}}}, \bibinfo{journal}{\apjl}
  \textbf{\bibinfo{volume}{819}}, \bibinfo{eid}{L17} (\bibinfo{year}{2016}),
  \eprint{1511.00005}.

\bibitem[{\citenamefont{{Stone} et~al.}(2017)\citenamefont{{Stone}, {Metzger},
  and {Haiman}}}]{StoneAGNI+2017}
\bibinfo{author}{\bibfnamefont{N.~C.} \bibnamefont{{Stone}}},
  \bibinfo{author}{\bibfnamefont{B.~D.} \bibnamefont{{Metzger}}},
  \bibnamefont{and} \bibinfo{author}{\bibfnamefont{Z.}~\bibnamefont{{Haiman}}},
  \bibinfo{journal}{\mnras} \textbf{\bibinfo{volume}{464}},
  \bibinfo{pages}{946} (\bibinfo{year}{2017}), \eprint{1602.04226}.

\bibitem[{\citenamefont{{Bartos} et~al.}(2017)\citenamefont{{Bartos}, {Kocsis},
  {Haiman}, and {M{\'a}rka}}}]{BartosAGNII+2017}
\bibinfo{author}{\bibfnamefont{I.}~\bibnamefont{{Bartos}}},
  \bibinfo{author}{\bibfnamefont{B.}~\bibnamefont{{Kocsis}}},
  \bibinfo{author}{\bibfnamefont{Z.}~\bibnamefont{{Haiman}}}, \bibnamefont{and}
  \bibinfo{author}{\bibfnamefont{S.}~\bibnamefont{{M{\'a}rka}}},
  \bibinfo{journal}{\apj} \textbf{\bibinfo{volume}{835}}, \bibinfo{eid}{165}
  (\bibinfo{year}{2017}), \eprint{1602.03831}.

\bibitem[{\citenamefont{{McKernan} et~al.}(2018)\citenamefont{{McKernan},
  {Ford}, {Bellovary}, {Leigh}, {Haiman}, {Kocsis}, {Lyra}, {Mac Low},
  {Metzger}, {O'Dowd} et~al.}}]{McKernanBBHAGN+2018}
\bibinfo{author}{\bibfnamefont{B.}~\bibnamefont{{McKernan}}},
  \bibinfo{author}{\bibfnamefont{K.~E.~S.} \bibnamefont{{Ford}}},
  \bibinfo{author}{\bibfnamefont{J.}~\bibnamefont{{Bellovary}}},
  \bibinfo{author}{\bibfnamefont{N.~W.~C.} \bibnamefont{{Leigh}}},
  \bibinfo{author}{\bibfnamefont{Z.}~\bibnamefont{{Haiman}}},
  \bibinfo{author}{\bibfnamefont{B.}~\bibnamefont{{Kocsis}}},
  \bibinfo{author}{\bibfnamefont{W.}~\bibnamefont{{Lyra}}},
  \bibinfo{author}{\bibfnamefont{M.~M.} \bibnamefont{{Mac Low}}},
  \bibinfo{author}{\bibfnamefont{B.}~\bibnamefont{{Metzger}}},
  \bibinfo{author}{\bibfnamefont{M.}~\bibnamefont{{O'Dowd}}},
  \bibnamefont{et~al.}, \bibinfo{journal}{\apj} \textbf{\bibinfo{volume}{866}},
  \bibinfo{eid}{66} (\bibinfo{year}{2018}), \eprint{1702.07818}.

\bibitem[{\citenamefont{{Amaro-Seoane} and et~al.}(2017)}]{LISA:2017}
\bibinfo{author}{\bibfnamefont{P.}~\bibnamefont{{Amaro-Seoane}}}
  \bibnamefont{and} \bibinfo{author}{\bibnamefont{et~al.}},
  \bibinfo{journal}{ArXiv e-prints}  (\bibinfo{year}{2017}),
  \eprint{1702.00786}.

\bibitem[{\citenamefont{{Chesler} and {Loeb}}(2017)}]{CheslerLoeb:2017}
\bibinfo{author}{\bibfnamefont{P.~M.} \bibnamefont{{Chesler}}}
  \bibnamefont{and} \bibinfo{author}{\bibfnamefont{A.}~\bibnamefont{{Loeb}}},
  \bibinfo{journal}{\prl} \textbf{\bibinfo{volume}{119}}, \bibinfo{eid}{031102}
  (\bibinfo{year}{2017}), \eprint{1704.05116}.

\bibitem[{\citenamefont{{Mardling} and {Aarseth}}(2001)}]{MardlingAraseth:2001}
\bibinfo{author}{\bibfnamefont{R.~A.} \bibnamefont{{Mardling}}}
  \bibnamefont{and} \bibinfo{author}{\bibfnamefont{S.~J.}
  \bibnamefont{{Aarseth}}}, \bibinfo{journal}{\mnras}
  \textbf{\bibinfo{volume}{321}}, \bibinfo{pages}{398} (\bibinfo{year}{2001}).

\bibitem[{\citenamefont{{Schneider} et~al.}(1992)\citenamefont{{Schneider},
  {Ehlers}, and {Falco}}}]{SEF:Glens:1992}
\bibinfo{author}{\bibfnamefont{P.}~\bibnamefont{{Schneider}}},
  \bibinfo{author}{\bibfnamefont{J.}~\bibnamefont{{Ehlers}}}, \bibnamefont{and}
  \bibinfo{author}{\bibfnamefont{E.~E.} \bibnamefont{{Falco}}},
  \emph{\bibinfo{title}{{Gravitational Lenses}}} (\bibinfo{year}{1992}).

\bibitem[{\citenamefont{{D'Orazio} and {Di Stefano}}(2018)}]{DoDi:2018}
\bibinfo{author}{\bibfnamefont{D.~J.} \bibnamefont{{D'Orazio}}}
  \bibnamefont{and} \bibinfo{author}{\bibfnamefont{R.}~\bibnamefont{{Di
  Stefano}}}, \bibinfo{journal}{\mnras} \textbf{\bibinfo{volume}{474}},
  \bibinfo{pages}{2975} (\bibinfo{year}{2018}), \eprint{1707.02335}.

\bibitem[{\citenamefont{{D'Orazio} and {Di Stefano}}(2019)}]{DoDi:2019}
\bibinfo{author}{\bibfnamefont{D.~J.} \bibnamefont{{D'Orazio}}}
  \bibnamefont{and} \bibinfo{author}{\bibfnamefont{R.}~\bibnamefont{{Di
  Stefano}}}, \bibinfo{journal}{arXiv e-prints} \bibinfo{eid}{arXiv:1906.11149}
  (\bibinfo{year}{2019}), \eprint{1906.11149}.

\bibitem[{\citenamefont{{Sesana}}(2016)}]{SesanaMultibandGW:2016}
\bibinfo{author}{\bibfnamefont{A.}~\bibnamefont{{Sesana}}},
  \bibinfo{journal}{\prl} \textbf{\bibinfo{volume}{116}}, \bibinfo{eid}{231102}
  (\bibinfo{year}{2016}), \eprint{1602.06951}.

\bibitem[{\citenamefont{{Robson} et~al.}(2019)\citenamefont{{Robson},
  {Cornish}, and {Liu}}}]{RobsonLISASens:2019}
\bibinfo{author}{\bibfnamefont{T.}~\bibnamefont{{Robson}}},
  \bibinfo{author}{\bibfnamefont{N.~J.} \bibnamefont{{Cornish}}},
  \bibnamefont{and} \bibinfo{author}{\bibfnamefont{C.}~\bibnamefont{{Liu}}},
  \bibinfo{journal}{Classical and Quantum Gravity}
  \textbf{\bibinfo{volume}{36}}, \bibinfo{eid}{105011} (\bibinfo{year}{2019}),
  \eprint{1803.01944}.

\bibitem[{\citenamefont{http://www.et gw.eu/}(2018)}]{ET}
\bibinfo{author}{\bibfnamefont{T.~E. T.~P.} \bibnamefont{http://www.et gw.eu/}}
  (\bibinfo{year}{2018}).

\bibitem[{\citenamefont{{Sesana} et~al.}(2005)\citenamefont{{Sesana}, {Haardt},
  {Madau}, and {Volonteri}}}]{Sesana+2005}
\bibinfo{author}{\bibfnamefont{A.}~\bibnamefont{{Sesana}}},
  \bibinfo{author}{\bibfnamefont{F.}~\bibnamefont{{Haardt}}},
  \bibinfo{author}{\bibfnamefont{P.}~\bibnamefont{{Madau}}}, \bibnamefont{and}
  \bibinfo{author}{\bibfnamefont{M.}~\bibnamefont{{Volonteri}}},
  \bibinfo{journal}{\apj} \textbf{\bibinfo{volume}{623}}, \bibinfo{pages}{23}
  (\bibinfo{year}{2005}), \eprint{astro-ph/0409255}.

\bibitem[{\citenamefont{{D'Orazio} and {Samsing}}(2018)}]{SD3:2018}
\bibinfo{author}{\bibfnamefont{D.~J.} \bibnamefont{{D'Orazio}}}
  \bibnamefont{and}
  \bibinfo{author}{\bibfnamefont{J.}~\bibnamefont{{Samsing}}},
  \bibinfo{journal}{\mnras} \textbf{\bibinfo{volume}{481}},
  \bibinfo{pages}{4775} (\bibinfo{year}{2018}), \eprint{1805.06194}.

\bibitem[{\citenamefont{{Bahcall} and {Wolf}}(1976)}]{BW76}
\bibinfo{author}{\bibfnamefont{J.~N.} \bibnamefont{{Bahcall}}}
  \bibnamefont{and} \bibinfo{author}{\bibfnamefont{R.~A.}
  \bibnamefont{{Wolf}}}, \bibinfo{journal}{\apj}
  \textbf{\bibinfo{volume}{209}}, \bibinfo{pages}{214} (\bibinfo{year}{1976}).

\bibitem[{\citenamefont{{Bahcall} and {Wolf}}(1977)}]{BW77}
\bibinfo{author}{\bibfnamefont{J.~N.} \bibnamefont{{Bahcall}}}
  \bibnamefont{and} \bibinfo{author}{\bibfnamefont{R.~A.}
  \bibnamefont{{Wolf}}}, \bibinfo{journal}{\apj}
  \textbf{\bibinfo{volume}{216}}, \bibinfo{pages}{883} (\bibinfo{year}{1977}).

\bibitem[{\citenamefont{{O'Leary} et~al.}(2009)\citenamefont{{O'Leary},
  {Kocsis}, and {Loeb}}}]{OlearyKocLoeb:2009}
\bibinfo{author}{\bibfnamefont{R.~M.} \bibnamefont{{O'Leary}}},
  \bibinfo{author}{\bibfnamefont{B.}~\bibnamefont{{Kocsis}}}, \bibnamefont{and}
  \bibinfo{author}{\bibfnamefont{A.}~\bibnamefont{{Loeb}}},
  \bibinfo{journal}{\mnras} \textbf{\bibinfo{volume}{395}},
  \bibinfo{pages}{2127} (\bibinfo{year}{2009}), \eprint{0807.2638}.

\bibitem[{\citenamefont{{Shankar} et~al.}(2004)\citenamefont{{Shankar},
  {Salucci}, {Granato}, {De Zotti}, and {Danese}}}]{Shankar:2004}
\bibinfo{author}{\bibfnamefont{F.}~\bibnamefont{{Shankar}}},
  \bibinfo{author}{\bibfnamefont{P.}~\bibnamefont{{Salucci}}},
  \bibinfo{author}{\bibfnamefont{G.~L.} \bibnamefont{{Granato}}},
  \bibinfo{author}{\bibfnamefont{G.}~\bibnamefont{{De Zotti}}},
  \bibnamefont{and} \bibinfo{author}{\bibfnamefont{L.}~\bibnamefont{{Danese}}},
  \bibinfo{journal}{\mnras} \textbf{\bibinfo{volume}{354}},
  \bibinfo{pages}{1020} (\bibinfo{year}{2004}), \eprint{astro-ph/0405585}.

\bibitem[{\citenamefont{{Samsing} et~al.}(2019)\citenamefont{{Samsing},
  {D'Orazio}, {Kremer}, {Rodriguez}, and {Askar}}}]{SamsingSS+2019}
\bibinfo{author}{\bibfnamefont{J.}~\bibnamefont{{Samsing}}},
  \bibinfo{author}{\bibfnamefont{D.~J.} \bibnamefont{{D'Orazio}}},
  \bibinfo{author}{\bibfnamefont{K.}~\bibnamefont{{Kremer}}},
  \bibinfo{author}{\bibfnamefont{C.~L.} \bibnamefont{{Rodriguez}}},
  \bibnamefont{and} \bibinfo{author}{\bibfnamefont{A.}~\bibnamefont{{Askar}}},
  \bibinfo{journal}{arXiv e-prints} \bibinfo{eid}{arXiv:1907.11231}
  (\bibinfo{year}{2019}), \eprint{1907.11231}.

\bibitem[{\citenamefont{{Chen} and {Amaro-Seoane}}(2017)}]{ChenAmaro:2017}
\bibinfo{author}{\bibfnamefont{X.}~\bibnamefont{{Chen}}} \bibnamefont{and}
  \bibinfo{author}{\bibfnamefont{P.}~\bibnamefont{{Amaro-Seoane}}},
  \bibinfo{journal}{\apjl} \textbf{\bibinfo{volume}{842}}, \bibinfo{eid}{L2}
  (\bibinfo{year}{2017}), \eprint{1702.08479}.

\bibitem[{\citenamefont{{Secunda} et~al.}(2018)\citenamefont{{Secunda},
  {Bellovary}, {Mac Low}, {Ford}, {McKernan}, {Leigh}, and
  {Lyra}}}]{Secunda+2018}
\bibinfo{author}{\bibfnamefont{A.}~\bibnamefont{{Secunda}}},
  \bibinfo{author}{\bibfnamefont{J.}~\bibnamefont{{Bellovary}}},
  \bibinfo{author}{\bibfnamefont{M.-M.} \bibnamefont{{Mac Low}}},
  \bibinfo{author}{\bibfnamefont{K.~E.~S.} \bibnamefont{{Ford}}},
  \bibinfo{author}{\bibfnamefont{B.}~\bibnamefont{{McKernan}}},
  \bibinfo{author}{\bibfnamefont{N.}~\bibnamefont{{Leigh}}}, \bibnamefont{and}
  \bibinfo{author}{\bibfnamefont{W.}~\bibnamefont{{Lyra}}},
  \bibinfo{journal}{arXiv e-prints}  (\bibinfo{year}{2018}),
  \eprint{1807.02859}.

\bibitem[{\citenamefont{{Yang} et~al.}(2019)\citenamefont{{Yang}, {Bartos},
  {Gayathri}, {Ford}, {Haiman}, {Klimenko}, {Kocsis}, {M{\'a}rka}, {M{\'a}rka},
  {McKernan} et~al.}}]{YangBartos+2019}
\bibinfo{author}{\bibfnamefont{Y.}~\bibnamefont{{Yang}}},
  \bibinfo{author}{\bibfnamefont{I.}~\bibnamefont{{Bartos}}},
  \bibinfo{author}{\bibfnamefont{V.}~\bibnamefont{{Gayathri}}},
  \bibinfo{author}{\bibfnamefont{S.}~\bibnamefont{{Ford}}},
  \bibinfo{author}{\bibfnamefont{Z.}~\bibnamefont{{Haiman}}},
  \bibinfo{author}{\bibfnamefont{S.}~\bibnamefont{{Klimenko}}},
  \bibinfo{author}{\bibfnamefont{B.}~\bibnamefont{{Kocsis}}},
  \bibinfo{author}{\bibfnamefont{S.}~\bibnamefont{{M{\'a}rka}}},
  \bibinfo{author}{\bibfnamefont{Z.}~\bibnamefont{{M{\'a}rka}}},
  \bibinfo{author}{\bibfnamefont{B.}~\bibnamefont{{McKernan}}},
  \bibnamefont{et~al.}, \bibinfo{journal}{arXiv e-prints}
  \bibinfo{eid}{arXiv:1906.09281} (\bibinfo{year}{2019}), \eprint{1906.09281}.

\bibitem[{\citenamefont{{Sirko} and {Goodman}}(2003)}]{SirkoGoodman:2003}
\bibinfo{author}{\bibfnamefont{E.}~\bibnamefont{{Sirko}}} \bibnamefont{and}
  \bibinfo{author}{\bibfnamefont{J.}~\bibnamefont{{Goodman}}},
  \bibinfo{journal}{\mnras} \textbf{\bibinfo{volume}{341}},
  \bibinfo{pages}{501} (\bibinfo{year}{2003}), \eprint{astro-ph/0209469}.

\bibitem[{\citenamefont{{Tang} et~al.}(2017)\citenamefont{{Tang}, {MacFadyen},
  and {Haiman}}}]{Tang+2017}
\bibinfo{author}{\bibfnamefont{Y.}~\bibnamefont{{Tang}}},
  \bibinfo{author}{\bibfnamefont{A.}~\bibnamefont{{MacFadyen}}},
  \bibnamefont{and} \bibinfo{author}{\bibfnamefont{Z.}~\bibnamefont{{Haiman}}},
  \bibinfo{journal}{\mnras} \textbf{\bibinfo{volume}{469}},
  \bibinfo{pages}{4258} (\bibinfo{year}{2017}), \eprint{1703.03913}.

\bibitem[{\citenamefont{{Mu{\~n}oz} et~al.}(2019)\citenamefont{{Mu{\~n}oz},
  {Miranda}, and {Lai}}}]{MunozLai+2019}
\bibinfo{author}{\bibfnamefont{D.~J.} \bibnamefont{{Mu{\~n}oz}}},
  \bibinfo{author}{\bibfnamefont{R.}~\bibnamefont{{Miranda}}},
  \bibnamefont{and} \bibinfo{author}{\bibfnamefont{D.}~\bibnamefont{{Lai}}},
  \bibinfo{journal}{\apj} \textbf{\bibinfo{volume}{871}}, \bibinfo{eid}{84}
  (\bibinfo{year}{2019}), \eprint{1810.04676}.

\bibitem[{\citenamefont{{Duffell} et~al.}(2019)\citenamefont{{Duffell},
  {D'Orazio}, {Derdzinski}, {Haiman}, {MacFadyen}, {Rosen}, and
  {Zrake}}}]{DuffellCBDq+2019}
\bibinfo{author}{\bibfnamefont{P.~C.} \bibnamefont{{Duffell}}},
  \bibinfo{author}{\bibfnamefont{D.}~\bibnamefont{{D'Orazio}}},
  \bibinfo{author}{\bibfnamefont{A.}~\bibnamefont{{Derdzinski}}},
  \bibinfo{author}{\bibfnamefont{Z.}~\bibnamefont{{Haiman}}},
  \bibinfo{author}{\bibfnamefont{A.}~\bibnamefont{{MacFadyen}}},
  \bibinfo{author}{\bibfnamefont{A.~L.} \bibnamefont{{Rosen}}},
  \bibnamefont{and} \bibinfo{author}{\bibfnamefont{J.}~\bibnamefont{{Zrake}}},
  \bibinfo{journal}{arXiv e-prints} \bibinfo{eid}{arXiv:1911.05506}
  (\bibinfo{year}{2019}), \eprint{1911.05506}.

\bibitem[{\citenamefont{{Kremer} et~al.}(2019)\citenamefont{{Kremer},
  {Rodriguez}, {Amaro-Seoane}, {Breivik}, {Chatterjee}, {Katz}, {Larson},
  {Rasio}, {Samsing}, {Ye} et~al.}}]{KremerLISA+2019}
\bibinfo{author}{\bibfnamefont{K.}~\bibnamefont{{Kremer}}},
  \bibinfo{author}{\bibfnamefont{C.~L.} \bibnamefont{{Rodriguez}}},
  \bibinfo{author}{\bibfnamefont{P.}~\bibnamefont{{Amaro-Seoane}}},
  \bibinfo{author}{\bibfnamefont{K.}~\bibnamefont{{Breivik}}},
  \bibinfo{author}{\bibfnamefont{S.}~\bibnamefont{{Chatterjee}}},
  \bibinfo{author}{\bibfnamefont{M.~L.} \bibnamefont{{Katz}}},
  \bibinfo{author}{\bibfnamefont{S.~L.} \bibnamefont{{Larson}}},
  \bibinfo{author}{\bibfnamefont{F.~A.} \bibnamefont{{Rasio}}},
  \bibinfo{author}{\bibfnamefont{J.}~\bibnamefont{{Samsing}}},
  \bibinfo{author}{\bibfnamefont{C.~S.} \bibnamefont{{Ye}}},
  \bibnamefont{et~al.}, \bibinfo{journal}{\prd} \textbf{\bibinfo{volume}{99}},
  \bibinfo{eid}{063003} (\bibinfo{year}{2019}), \eprint{1811.11812}.

\bibitem[{\citenamefont{{Gerosa} et~al.}(2019)\citenamefont{{Gerosa}, {Ma},
  {Wong}, {Berti}, {O'Shaughnessy}, {Chen}, and {Belczynski}}}]{Gerosa+2019}
\bibinfo{author}{\bibfnamefont{D.}~\bibnamefont{{Gerosa}}},
  \bibinfo{author}{\bibfnamefont{S.}~\bibnamefont{{Ma}}},
  \bibinfo{author}{\bibfnamefont{K.~W.~K.} \bibnamefont{{Wong}}},
  \bibinfo{author}{\bibfnamefont{E.}~\bibnamefont{{Berti}}},
  \bibinfo{author}{\bibfnamefont{R.}~\bibnamefont{{O'Shaughnessy}}},
  \bibinfo{author}{\bibfnamefont{Y.}~\bibnamefont{{Chen}}}, \bibnamefont{and}
  \bibinfo{author}{\bibfnamefont{K.}~\bibnamefont{{Belczynski}}},
  \bibinfo{journal}{\prd} \textbf{\bibinfo{volume}{99}}, \bibinfo{eid}{103004}
  (\bibinfo{year}{2019}), \eprint{1902.00021}.

\bibitem[{\citenamefont{{Nishizawa} et~al.}(2017)\citenamefont{{Nishizawa},
  {Sesana}, {Berti}, and {Klein}}}]{Nishizawa+2017}
\bibinfo{author}{\bibfnamefont{A.}~\bibnamefont{{Nishizawa}}},
  \bibinfo{author}{\bibfnamefont{A.}~\bibnamefont{{Sesana}}},
  \bibinfo{author}{\bibfnamefont{E.}~\bibnamefont{{Berti}}}, \bibnamefont{and}
  \bibinfo{author}{\bibfnamefont{A.}~\bibnamefont{{Klein}}},
  \bibinfo{journal}{\mnras} \textbf{\bibinfo{volume}{465}},
  \bibinfo{pages}{4375} (\bibinfo{year}{2017}), \eprint{1606.09295}.

\bibitem[{\citenamefont{{Randall} and
  {Xianyu}}(2018)}]{RandallZhongZhiEcc:2018}
\bibinfo{author}{\bibfnamefont{L.}~\bibnamefont{{Randall}}} \bibnamefont{and}
  \bibinfo{author}{\bibfnamefont{Z.-Z.} \bibnamefont{{Xianyu}}},
  \bibinfo{journal}{\apj} \textbf{\bibinfo{volume}{853}}, \bibinfo{eid}{93}
  (\bibinfo{year}{2018}), \eprint{1708.08569}.

\bibitem[{\citenamefont{{Fragione} et~al.}(2019)\citenamefont{{Fragione},
  {Grishin}, {Leigh}, {Perets}, and {Perna}}}]{GiacomoKL+2019}
\bibinfo{author}{\bibfnamefont{G.}~\bibnamefont{{Fragione}}},
  \bibinfo{author}{\bibfnamefont{E.}~\bibnamefont{{Grishin}}},
  \bibinfo{author}{\bibfnamefont{N.~W.~C.} \bibnamefont{{Leigh}}},
  \bibinfo{author}{\bibfnamefont{H.~B.} \bibnamefont{{Perets}}},
  \bibnamefont{and} \bibinfo{author}{\bibfnamefont{R.}~\bibnamefont{{Perna}}},
  \bibinfo{journal}{\mnras} \textbf{\bibinfo{volume}{488}}, \bibinfo{pages}{47}
  (\bibinfo{year}{2019}), \eprint{1811.10627}.

\bibitem[{\citenamefont{{Samsing}}(2018)}]{Samsing:2018}
\bibinfo{author}{\bibfnamefont{J.}~\bibnamefont{{Samsing}}},
  \bibinfo{journal}{\prd} \textbf{\bibinfo{volume}{97}}, \bibinfo{eid}{103014}
  (\bibinfo{year}{2018}), \eprint{1711.07452}.

\bibitem[{\citenamefont{{Rodriguez} et~al.}(2018)\citenamefont{{Rodriguez},
  {Amaro-Seoane}, {Chatterjee}, {Kremer}, {Rasio}, {Samsing}, {Ye}, and
  {Zevin}}}]{RodPNCMC+2018}
\bibinfo{author}{\bibfnamefont{C.~L.} \bibnamefont{{Rodriguez}}},
  \bibinfo{author}{\bibfnamefont{P.}~\bibnamefont{{Amaro-Seoane}}},
  \bibinfo{author}{\bibfnamefont{S.}~\bibnamefont{{Chatterjee}}},
  \bibinfo{author}{\bibfnamefont{K.}~\bibnamefont{{Kremer}}},
  \bibinfo{author}{\bibfnamefont{F.~A.} \bibnamefont{{Rasio}}},
  \bibinfo{author}{\bibfnamefont{J.}~\bibnamefont{{Samsing}}},
  \bibinfo{author}{\bibfnamefont{C.~S.} \bibnamefont{{Ye}}}, \bibnamefont{and}
  \bibinfo{author}{\bibfnamefont{M.}~\bibnamefont{{Zevin}}},
  \bibinfo{journal}{\prd} \textbf{\bibinfo{volume}{98}}, \bibinfo{eid}{123005}
  (\bibinfo{year}{2018}), \eprint{1811.04926}.

\bibitem[{\citenamefont{{Samsing} and {D'Orazio}}(2018)}]{SD2:2018}
\bibinfo{author}{\bibfnamefont{J.}~\bibnamefont{{Samsing}}} \bibnamefont{and}
  \bibinfo{author}{\bibfnamefont{D.~J.} \bibnamefont{{D'Orazio}}},
  \bibinfo{journal}{\mnras} \textbf{\bibinfo{volume}{481}},
  \bibinfo{pages}{5445} (\bibinfo{year}{2018}), \eprint{1804.06519}.

\bibitem[{\citenamefont{{Hu} et~al.}(2019)\citenamefont{{Hu}, {D'Orazio},
  {Haiman}, {Smith}, {Snios}, {Charisi}, and {Di Stefano}}}]{HuSpikey+2019}
\bibinfo{author}{\bibfnamefont{B.~X.} \bibnamefont{{Hu}}},
  \bibinfo{author}{\bibfnamefont{D.~J.} \bibnamefont{{D'Orazio}}},
  \bibinfo{author}{\bibfnamefont{Z.}~\bibnamefont{{Haiman}}},
  \bibinfo{author}{\bibfnamefont{K.~L.} \bibnamefont{{Smith}}},
  \bibinfo{author}{\bibfnamefont{B.}~\bibnamefont{{Snios}}},
  \bibinfo{author}{\bibfnamefont{M.}~\bibnamefont{{Charisi}}},
  \bibnamefont{and} \bibinfo{author}{\bibfnamefont{R.}~\bibnamefont{{Di
  Stefano}}}, \bibinfo{journal}{arXiv e-prints} \bibinfo{eid}{arXiv:1910.05348}
  (\bibinfo{year}{2019}), \eprint{1910.05348}.

\bibitem[{\citenamefont{{Liu} et~al.}(2019)\citenamefont{{Liu}, {Lai}, and
  {Wang}}}]{BinLiuGRKL+2019}
\bibinfo{author}{\bibfnamefont{B.}~\bibnamefont{{Liu}}},
  \bibinfo{author}{\bibfnamefont{D.}~\bibnamefont{{Lai}}}, \bibnamefont{and}
  \bibinfo{author}{\bibfnamefont{Y.-H.} \bibnamefont{{Wang}}},
  \bibinfo{journal}{\apjl} \textbf{\bibinfo{volume}{883}}, \bibinfo{eid}{L7}
  (\bibinfo{year}{2019}), \eprint{1906.07726}.

\bibitem[{\citenamefont{{Jani} et~al.}(2019)\citenamefont{{Jani}, {Shoemaker},
  and {Cutler}}}]{JaniCutler:2019}
\bibinfo{author}{\bibfnamefont{K.}~\bibnamefont{{Jani}}},
  \bibinfo{author}{\bibfnamefont{D.}~\bibnamefont{{Shoemaker}}},
  \bibnamefont{and} \bibinfo{author}{\bibfnamefont{C.}~\bibnamefont{{Cutler}}},
  \bibinfo{journal}{arXiv e-prints} \bibinfo{eid}{arXiv:1908.04985}
  (\bibinfo{year}{2019}), \eprint{1908.04985}.

\bibitem[{\citenamefont{{Ebisuzaki} et~al.}(2001)\citenamefont{{Ebisuzaki},
  {Makino}, {Tsuru}, {Funato}, {Portegies Zwart}, {Hut}, {McMillan},
  {Matsushita}, {Matsumoto}, and {Kawabe}}}]{Ebisuzaki+2001}
\bibinfo{author}{\bibfnamefont{T.}~\bibnamefont{{Ebisuzaki}}},
  \bibinfo{author}{\bibfnamefont{J.}~\bibnamefont{{Makino}}},
  \bibinfo{author}{\bibfnamefont{T.~G.} \bibnamefont{{Tsuru}}},
  \bibinfo{author}{\bibfnamefont{Y.}~\bibnamefont{{Funato}}},
  \bibinfo{author}{\bibfnamefont{S.}~\bibnamefont{{Portegies Zwart}}},
  \bibinfo{author}{\bibfnamefont{P.}~\bibnamefont{{Hut}}},
  \bibinfo{author}{\bibfnamefont{S.}~\bibnamefont{{McMillan}}},
  \bibinfo{author}{\bibfnamefont{S.}~\bibnamefont{{Matsushita}}},
  \bibinfo{author}{\bibfnamefont{H.}~\bibnamefont{{Matsumoto}}},
  \bibnamefont{and} \bibinfo{author}{\bibfnamefont{R.}~\bibnamefont{{Kawabe}}},
  \bibinfo{journal}{\apjl} \textbf{\bibinfo{volume}{562}}, \bibinfo{pages}{L19}
  (\bibinfo{year}{2001}), \eprint{astro-ph/0106252}.

\bibitem[{\citenamefont{{Fragione} et~al.}(2018)\citenamefont{{Fragione},
  {Ginsburg}, and {Kocsis}}}]{FragioneIMBHa:2018}
\bibinfo{author}{\bibfnamefont{G.}~\bibnamefont{{Fragione}}},
  \bibinfo{author}{\bibfnamefont{I.}~\bibnamefont{{Ginsburg}}},
  \bibnamefont{and} \bibinfo{author}{\bibfnamefont{B.}~\bibnamefont{{Kocsis}}},
  \bibinfo{journal}{\apj} \textbf{\bibinfo{volume}{856}}, \bibinfo{eid}{92}
  (\bibinfo{year}{2018}), \eprint{1711.00483}.

\bibitem[{\citenamefont{{Kremer} et~al.}(2018)\citenamefont{{Kremer},
  {Chatterjee}, {Breivik}, {Rodriguez}, {Larson}, and
  {Rasio}}}]{KremerMWLISA+2018}
\bibinfo{author}{\bibfnamefont{K.}~\bibnamefont{{Kremer}}},
  \bibinfo{author}{\bibfnamefont{S.}~\bibnamefont{{Chatterjee}}},
  \bibinfo{author}{\bibfnamefont{K.}~\bibnamefont{{Breivik}}},
  \bibinfo{author}{\bibfnamefont{C.~L.} \bibnamefont{{Rodriguez}}},
  \bibinfo{author}{\bibfnamefont{S.~L.} \bibnamefont{{Larson}}},
  \bibnamefont{and} \bibinfo{author}{\bibfnamefont{F.~A.}
  \bibnamefont{{Rasio}}}, \bibinfo{journal}{\prl}
  \textbf{\bibinfo{volume}{120}}, \bibinfo{eid}{191103} (\bibinfo{year}{2018}),
  \eprint{1802.05661}.

\bibitem[{\citenamefont{{Nishizawa}}(2018)}]{NishizawaGWprop:2018}
\bibinfo{author}{\bibfnamefont{A.}~\bibnamefont{{Nishizawa}}},
  \bibinfo{journal}{\prd} \textbf{\bibinfo{volume}{97}}, \bibinfo{eid}{104037}
  (\bibinfo{year}{2018}), \eprint{1710.04825}.

\bibitem[{\citenamefont{{Haiman}}(2017)}]{HaimanEMChirp:2017}
\bibinfo{author}{\bibfnamefont{Z.}~\bibnamefont{{Haiman}}},
  \bibinfo{journal}{\prd} \textbf{\bibinfo{volume}{96}}, \bibinfo{eid}{023004}
  (\bibinfo{year}{2017}), \eprint{1705.06765}.

\bibitem[{\citenamefont{{Torres-Orjuela}
  et~al.}(2018)\citenamefont{{Torres-Orjuela}, {Chen}, {Cao}, {Amaro-Seoane},
  and {Peng}}}]{Torres-Orjuela+2018}
\bibinfo{author}{\bibfnamefont{A.}~\bibnamefont{{Torres-Orjuela}}},
  \bibinfo{author}{\bibfnamefont{X.}~\bibnamefont{{Chen}}},
  \bibinfo{author}{\bibfnamefont{Z.}~\bibnamefont{{Cao}}},
  \bibinfo{author}{\bibfnamefont{P.}~\bibnamefont{{Amaro-Seoane}}},
  \bibnamefont{and} \bibinfo{author}{\bibfnamefont{P.}~\bibnamefont{{Peng}}},
  \bibinfo{journal}{arXiv e-prints}  (\bibinfo{year}{2018}),
  \eprint{1806.09857}.

\end{thebibliography}
\end{document}